\documentclass[12pt]{iopart}

\newcommand{\gene}{\textsc{Gene} }

\newcommand{\kyrhos}{$k_y\rho_s$}
\newcommand{\ky}{{k_\perp\rho_s}}

\usepackage{graphics}
\usepackage{graphicx}
\usepackage{enumerate}
\usepackage{stfloats}

\begin{document}

\title[Measured and predicted turbulence frequency spectra]{Comparison between measured and predicted turbulence frequency spectra in ITG and TEM regimes}

\author{J.~Citrin$^{1,2}$,  H. Arnichand$^{2}$, J. Bernardo$^3$, C.~Bourdelle$^2$, X.~Garbet$^2$, F.~Jenko$^4$ S.~Hacquin$^2$, M.J. Pueschel$^5$, R. Sabot$^2$}

%\address{JET-EFDA, Culham Science Centre, Abingdon, OX14 3DB, UK}
\address{$^1$ DIFFER -- Dutch Institute for Fundamental Energy Research, De Zaale 20, 5612 AJ Eindhoven, the Netherlands}
\address{$^2$CEA, IRFM, F-13108 Saint Paul Lez Durance, France}
\address{$^3$Instituto de Plasmas e Fus\~ao Nuclear, Instituto Superior T\'ecnico, Universidade T\'ecnica de Lisboa, 1049-001 Lisboa, Portugal}
\address{$^4$Department of Physics and Astronomy, University of California, Los Angeles, California 90095, USA}
\address{$^5$Department of Physics, University of Wisconsin-Madison, Madison, Wisconsin 53706, USA}
\ead{J.Citrin@differ.nl}

\begin{abstract}
The observation of distinct peaks in tokamak core reflectometry measurements - named quasi-coherent-modes (QCMs) - are identified as a signature of Trapped-Electron-Mode (TEM) turbulence [H. Arnichand \textit{et al.} 2016 \textit{Plasma Phys. Control. Fusion} \textbf{58} 014037]. This phenomenon is investigated with detailed linear and nonlinear gyrokinetic simulations using the \gene code. A Tore-Supra density scan is studied, which traverses through a Linear (LOC) to Saturated (SOC) Ohmic Confinement transition. The LOC and SOC phases are both simulated separately. In the LOC phase, where QCMs are observed, TEMs are robustly predicted unstable in linear studies. In the later SOC phase, where QCMs are no longer observed, ITG modes are identified. In nonlinear simulations, in the ITG (SOC) phase, a broadband spectrum is seen. In the TEM (LOC) phase, a clear emergence of a peak at the TEM frequencies is seen. This is due to reduced nonlinear frequency broadening of the underlying linear modes in the TEM regime compared with the ITG regime. A synthetic diagnostic of the nonlinearly simulated frequency spectra reproduces the features observed in the reflectometry measurements. These results support the identification of core QCMs as an experimental marker for TEM turbulence.
\end{abstract}
%\pacs{1315, 9440T}
%\keywords{magnetic moment, solar neutrinos, astrophysics}

\section{Introduction}
The transport of particles, momentum, and heat in the tokamak core is dominated by turbulence driven by plasma microinstabilities over a range of spatiotemporal scales~\cite{hort99,ITER2}. At the ion scale, characterised by $\ky\approx[0.1-1]$, the most ubiquitous instabilities are modes driven by ion-temperature-gradients (ITG) and trapped-electron density and temperature gradients (TEM). $k_\perp$ is the perpendicular (to the magnetic field) wavenumber, and $\rho_{s}=\frac{\sqrt{T_em_i}}{Z_iB}$ is the main ion Larmor radius with respect to the main ion sound speed. Electron scale transport at $\ky\approx[6-60]$, driven unstable by electron-temperature-gradients (ETG), can further contribute to the electron heat flux, particularly in regimes when ion-scale-transport is weakened~\cite{jenk02,howa16}. 

The nature of the underlying modes which set the turbulence regime has profound implications for both total energy confinement as well as the profiles of the individual kinetic profiles. An ITG (TEM) regime leads to increased (decreased) electron density peaking with the application of core electron heating, with maximum peaking at the transition between the regimes~\cite{garb03,fabl10,angi12}. Heavy impurity transport is also strongly affected by the impact of the turbulence regime on the main ion density profile. This is through the neoclassical inward pinch driven by the density gradient, enhanced by poloidal asymmetries due to heating and rotation~\cite{angi14,cass15}. Rotation itself depends on the nature of the underlying modes~\cite{peet11,angi11,angi12}. The turbulence regime also impacts both heavy and light impurity transport directly~\cite{vill10,cass13,angi15}.

The study of the mechanisms and characteristics of the various turbulence regimes is thus of high importance. The underlying modes and resulting turbulence are described by the gyrokinetic model~\cite{garb10}. Experimental validation of the gyrokinetic predictions is key to increase trust in the fundamental model and  confidence in extrapolations to future devices. 

Validation can be carried out over a hierarchy of spatiotemporal scales. At a more coarse level, this consists of comparing predicted heat and particle fluxes to the experimental values derived from source calculations and power balance. Due to advances in increasing computer power these comparisons are increasing in physics complexity, with increasingly routine agreement, e.g.~\cite{goer14,citr15b,nava16,howa16}.  More detailed comparisons of fluctuation characteristics depend on microwave diagnostics, such as reflectometry (for electron density) and electron cyclotron emission (for electron temperature). Quantities compared include normalized fluctuation amplitudes, wavenumber spectra, frequency spectra, cross-phases, and radial electric field~\cite{holl09,casaphd,casa09b,whit10,rhod11,whit13,stro15}. The identification of validation metrics is a key component of a robust, quantitative study~\cite{terr08,gree10,holl16}. However, in this work, the radial location of the reflectometry measurements ($\rho=0.18$) do not perfectly coincide with the radial location of the gyrokinetic simulations and simulated frequency spectra ($\rho=0.36$) due to the poor quality of $T_i$ charge exchange measurements at the more inner radii. This limitation is discussed further in section~\ref{sec:discharge}. The central comparisons are of a more qualitative nature and such metrics are not employed. $\rho$ is the normalized toroidal flux coordinate.

ITG and TEM instabilities are at the same spatial scale. This complicates their direct measurement with fluctuation diagnostics. The turbulence regime can be inferred from linear gyrokinetic predictions based on profile measurements. However, a direct experimental marker of the nature of the underlying modes would provide both corroboration of the gyrokinetic model, as well as valuable realtime information on the discharge characteristics. One method is the measurement of the turbulence rotation velocity with Doppler reflectometry~\cite{conw04}, sensitive to the turbulence regime due to the opposite signs of ITG and TEM phase velocities. However, a robust identification of the underlying modes demands determination of the radial electric field, since $V_{E{\times}B}$ typically dominates over the mode phase velocity~\cite{conw06}. This is prone to uncertainties. 

A robust, direct marker of TEM turbulence has been recently identified. These are distinct peaks, named ``quasi-coherent modes'' (QCM) appearing in frequency spectrum reflectometry measurements in a wide range of tokamak core regions. QCMs have intermediate bandwidth (10s of kHz), and emerge above the broadband (100s of kHz) frequency spectrum~\cite{arni14,arni15,arni16,arniphd,erns16,zhon16}. They have been observed in Tore-Supra, T-10, TEXTOR, JET, DIII-D, HL2A and TEXT-U. Detailed linear and nonlinear gyrokinetic simulations of a subset of these discharges show that QCMs appear in conjunction with predicted TEM turbulence. In line with Ref.~\cite{arni15,arni16}, we refer to these observations henceforth as QC-TEM. In regimes with ITG dominated turbulence, the reflectometry spectrum remains broadband. 

We note that in parallel, there are observations of modes with similar quasi-coherent spectral signatures in the vicinity of edge transport barriers in multiple devices and regimes~\cite{dial14,wang14,golf14,terr14}. These modes are not identified with TEMs, and their provenance is under active investigation. Therefore, QC-TEMs do not unify all quasi-coherent mode observations, and are limited to the observations in the plasma core region inside the pedestal.

This paper focuses on the linear and nonlinear gyrokinetic simulations of ITG and TEM regimes in Tore-Supra Ohmic discharge 48102, using the \gene code~\cite{jenk00}. Nonlinear local gyrokinetic simulations in conjunction with a reflectometry synthetic diagnostic~\cite{hacq16} reproduces the observed behaviour; a broadband spectrum is observed in the ITG regime and QC-TEM in the TEM regime. While brief summaries of these results were presented in Refs.~\cite{arni14,arni16}, this present publication goes into greater depth. This includes sensitivity studies, and extended details of the frequency spectrum characteristics. 

Beyond the direct comparison with experimental observations, this work also relates to the more general question of mode frequency spectrum broadening in nonlinear simulations. Increased understanding of this phenomenon is important for the robust construction of reduced transport models in wide regions of parameter space, also where ITG modes and TEMs are concurrent~\cite{merz10}. Qualitative differences in frequency broadening is shown to impact the quasilinear saturation rules~\cite{citr12}. 
 
The rest of this paper is organized as follows. Section 2 details the experimental discharge and input parameters, section 3 outlines the modelling methods used, section 4 describes the linear gyrokinetic simulations, and section 5 describes the nonlinear gyrokinetic simulations and synthetic diagnostic results. Conclusions are made in section 6. 

\section{Experimental discharge}
\label{sec:discharge}
\begin{figure}[htbp]
		\centering
		\includegraphics[scale=0.8]{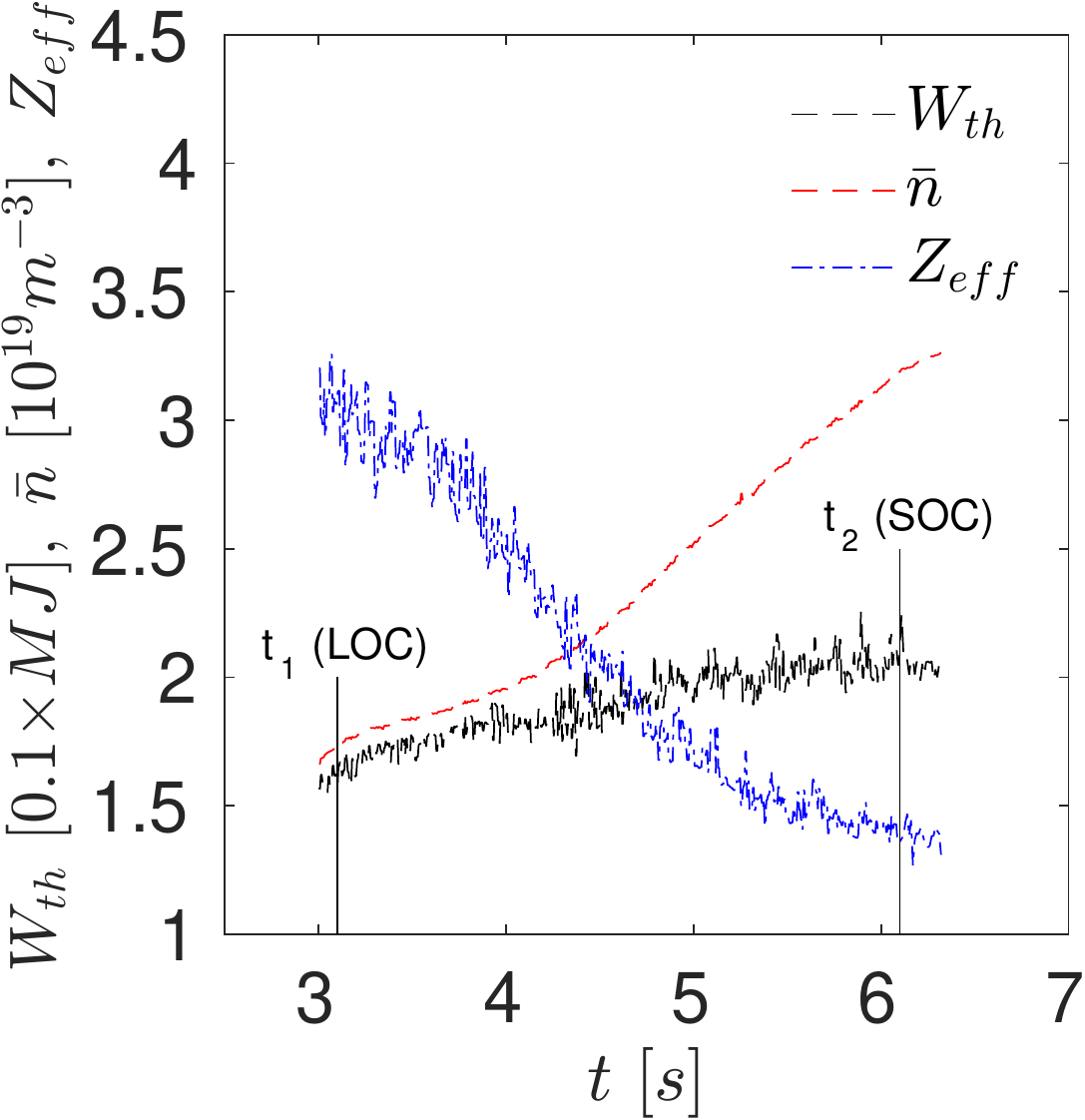}
		\caption{\footnotesize Stored thermal energy, line averaged density, and $Z_{eff}$ (from Bremsstrahlung) evolution of TS 48102 during current flattop ($I_p=1~\mathrm{MA}$). $t_1\approx3.1~s$ is in a LOC phase, and $t_2\approx6.1~s$ is in a SOC phase. These times are taken for the reflectometry measurements and gyrokinetic analysis. The saturated confinement is seen by the saturation of the stored thermal energy with increasing $\bar{n}$}
	\label{fig:figure1}
\end{figure}

The discharge selected for analysis is Tore Supra Ohmic discharge 48102. It is characterized by a density ramp achieved through gas puffing, and traverses through a SOC-LOC transition. The details of the discharge parameters and measurement configuration is extensively outlined in Ref.~\cite{bern15}, and for brevity not repeated here. Only a basic overview of the density rampup during current flattop is displayed in figure~\ref{fig:figure1}. In the plot, $t_1$ corresponds to a timeslice in the LOC regime (TEM turbulence) and $t_2$ to a timeslice in the SOC regime (ITG turbulence). The main impurity is carbon, and all simulations in this paper include both main ions and the C impurity.

\begin{figure}[htbp]
	\centering
	\includegraphics[scale=0.7]{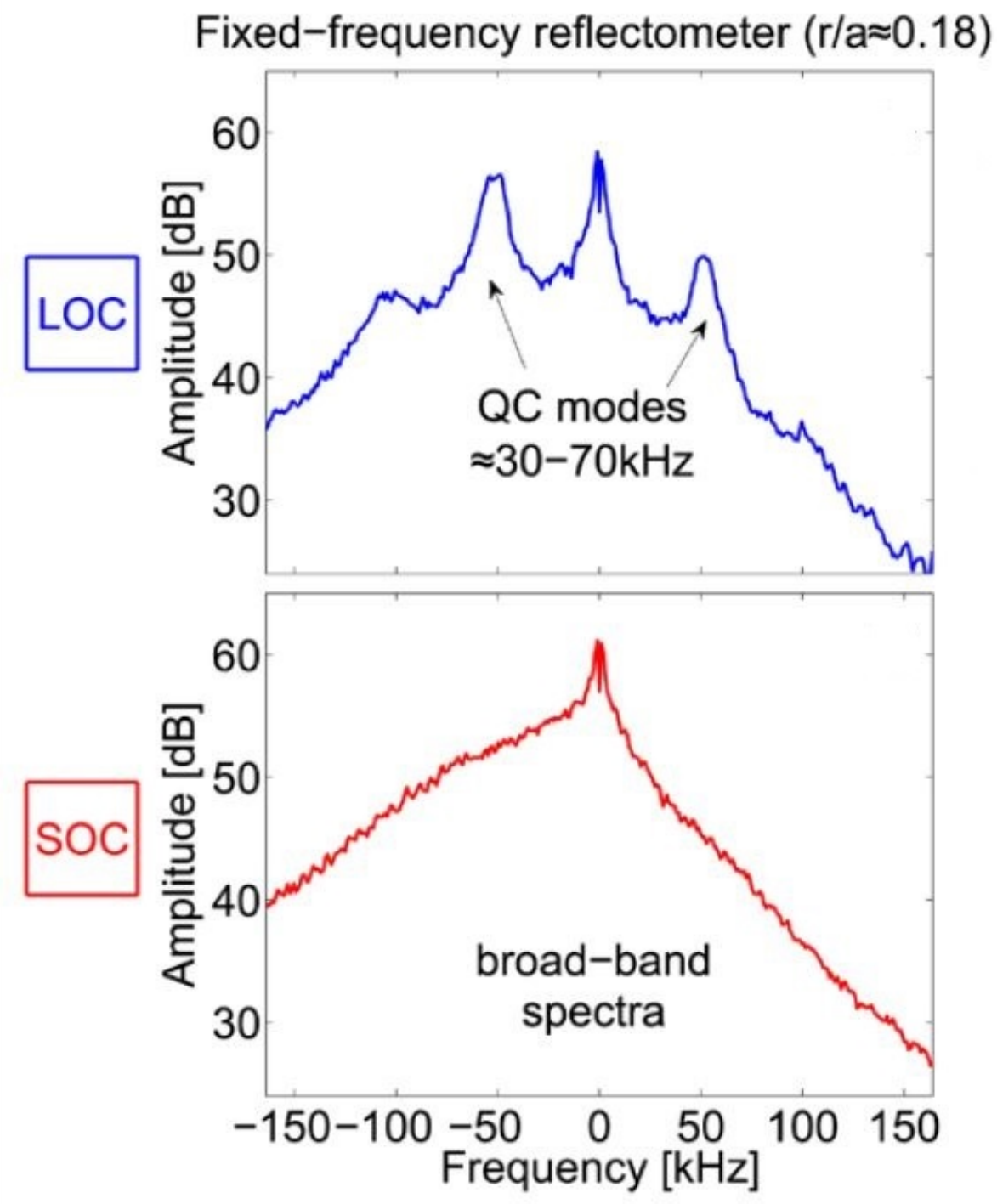}
	\caption{\footnotesize Fluctuation spectra from Tore Supra discharge 48102 at the $t_1$ LOC phase (upper panel) showing quasi-coherent-modes, and at the $t_2$ SOC phase (lower panel), showing a broadband spectrum. Figure reproduced with permission from Ref.~\cite{arni16}.}
	\label{fig:figure2}
\end{figure}

The turbulence measured in the SOC phase is characterised by a broadband frequency spectrum. However, in the LOC phase, distinct peaks can be distinguished within the spectrum. See figure~\ref{fig:figure2}, where the distinct peaks (quasi-coherent-modes) appear in the 50~kHz range. The provenance of these peaks is due to TEMs, according to gyrokinetic simulations. Note that in both cases, the central peak is not related to the turbulence, and is rather related to the intensity of the carrier wave of the reflectometer.

The dimensionless parameters extracted from the measurements are shown in Table 1. These are used as input for the gyrokinetic simulations. They correspond to profile averages over a $0.1~s$ window centred around $t_1$ and $t_2$ respectively. $Z_{eff}$ is due to a carbon impurity species. The carbon temperature and density gradients are set identical to the main ion, unless otherwise specified.

A caveat is attached to the radial location of the simulated data compared to the measured data. The frequency spectra measurement location was constrained to $\rho=0.18$, set by tuning the frequency source to a specific density and magnetic field. However, due to limitations in the kinetic profile diagnostics at inner radii – particularly, but not limited to, charge-exchange $T_i$ measurements - robust density and temperature gradients were ultimately not available at this radial location. Hence it was not possible to precisely deduce the input parameters necessary for the local turbulence simulation at this $\rho=0.18$ position. A choice was thus made to carry out the simulation at $\rho=0.37$ where the kinetic profile gradients could be measured with higher precision. Our comparison is thus not strictly quantitative. However, the relevance of the qualitative comparison is valid; QC-TEMs have been observed at a wide range of radii in the regimes where they appear~\cite{arni15}. Furthermore, it is reasonable, at least for the Ohmic cases, that the change of turbulence regime responsible for the QC-TEM appears in a wide band of radii. The LOC-SOC transition is defined with respect to confinement time trends, indicating an impact on the total stored energy which is a volume integrated quantity. Since the LOC-SOC transition is linked with a change in turbulence regime, it is thus likely that the change in turbulence regime encompasses a significant volume of the plasma core. Furthermore, as discussed is section~\ref{sec:lin}, the main reason for the TEM to ITG transition is a reduction in main ion dilution due to impurity flushing at higher density. This occurs at all radii. For all the reasons listed above, we deem that any qualitative trend in predicted turbulence spectra at $\rho=0.37$ in this discharge, has relevance to the physical mechanism behind the observed trends at $\rho=0.18$.

While we focus the analysis at more inner core radii, we postulate that QC modes are markers of TEM turbulence when observed at any radius within the pedestal-forming-region or pedestal itself if in H-mode ($\rho<\sim0.95$). The only electrostatic ion-scale microturbulence expected to be active throughout this region is ITG or TEM, since collisionality is typically still too high for collisional modes, such as of the resistive ballooning type~\cite{bour12}. Furthermore, due to the lower $\beta^\prime$ at higher radii such as at $\rho\sim0.8$, (i.e. while still within the pedestal), it is extremely unlikely that other modes, such as electromagnetic Kinetic Ballooning Modes (KBM), are present there, particularly if not observed (as in our case) at the more inner radii where $\beta^\prime$ is higher. Thus, the ion-scale ITG-TEM paradigm should hold for this wide range of radii. In the very edge region however, $\rho>\sim0.95$, other modes may be the source of QC fluctuations observed there.

\begin{table*}[tp]
\small
\centering
\caption{\footnotesize Tore Supra 48102 dimensionless parameters as input into the local $\textsc{Gene}$ simulations, at $\rho=0.37$, where $\rho$ is the normalized toroidal flux coordinate. The profiles are averaged over the time window specified in the first column. The gradient lengths are defined taking the radial coordinate as the toroidal flux coordinate. The safety factor $q$ and the magnetic shear $\hat{s}$ are calculated from the CRONOS~\cite{arta10} interpretative simulation, and do not include any statistical errors. $\nu^*$ is the normalized electron collisionality: $\nu^*{\equiv}\nu_{ei}\frac{qR}{\epsilon^{1.5}v_{te}}$, with $\epsilon=a/R$ and $v_\mathrm{te}\equiv\sqrt{T_e/m_e}$. $Z_{\mathrm{eff}}$ is from Bremsstrahlung measurements, 
and a flat $Z_{\mathrm{eff}}$ profile is assumed. All error bars quoted are statistical errors, meaning the variance of the measured fluctuations of each given quantity over the time-window averaged over for each phase.  We cannot distinguish between physical fluctuations and measurement errors. We do not take into consideration potential systematic errors. However, since both phases were taken from the same discharge, any systematic errors are likely the same for both cases, preserving the observed differences observed between the cases. Unless otherwise specified (due to sensitivity tests), the impurity (carbon) logarithmic temperature gradient is taken identical to the main ion temperature gradient, assuming thermal equilibrium between ion species. Due to a lack of available impurity density profiles, the logarithmic density gradient of all species is assumed equal}
%\vspace{0.15cm}
\tabcolsep=0.11cm
\scalebox{0.9}{\begin{tabular}{c|c|c|c|c|c|c|c|c|c}
\label{tab:summary}
Phase & $R/L_{Ti}$ & $R/L_{Te}$ & $R/L_{n}$ & $T_e/T_i$ & $\beta_e$ [\%] & $\hat{s}$ & $q$ & $\nu^*$ & $Z_{\mathrm{eff}}$ \\
\hline
LOC (t=3.05-3.15 s) & 4.3$\pm$0.5 & 9.2$\pm$0.35 & 2.8$\pm$0.1 & 1.8$\pm$0.1 & 0.13 & 0.7 & 1.3 & 0.012 & 3.0$\pm$0.1 \\
SOC (t=6.05-6.15 s) & 5.0$\pm$0.4 & 8.9$\pm$0.25 & 1.8$\pm$0.1 & 1.6$\pm$0.1 & 0.14 & 0.75 & 1.25 & 0.029 & 1.4$\pm$0.1 \\
\end{tabular}}
\end{table*}

\section{Numerical methods}
In this section, the numerical tools applied for analysis of the discharge are briefly outlined.
\subsection{Gyrokinetic simulations}
The local, gradient-driven, ${\delta}f$ version of the \gene code is employed throughout this work. Both linear (with initial value solver) and nonlinear simulations were carried out. \gene is a Eulerian gyrokinetic code, evolving the perturbed particle distribution functions self-consistently with the Maxwell field equations. \gene works in field line coordinates, where $x$ is the radial coordinate, $z$ is the parallel coordinate along the field line, and $y$ is the binormal coordinate. The simulations were spectral in both the $x$ and $y$ directions.

Numerical, non-parametrized geometry was applied. The geometric coefficients were calculated by the \textsc{Tracer-Efit}~\cite{xanth06,toldphd} module in \gene, reading \textit{Eqdsk} data derived from CRONOS~\cite{arta10} interpretative integrated modelling simulations of the Tore-Supra discharge 48102. In CRONOS, the Grad-Shafranov equation was solved by the HELENA~\cite{huys91} code. Collisions in \gene were modelled using a linearised Landau-–Boltzmann operator. Typical grid parameters in the nonlinear simulation were as follows: perpendicular box sizes $[L_x,L_y]\approx[110, 125]$ in units of ion Larmor radii, perpendicular grid discretisations $[n_{kx}, n_{ky}]=[128,32]$, 24 point discretisation in the parallel direction, 48 points in the parallel velocity direction and 12 magnetic moments. The parallel velocity box ranged between $[-3v_{Tj},+3v_{Tj}]$, with thermal velocity $v_{Tj}\equiv\sqrt{2T_{j0}/m_j}$, where $T_{j0}$ is the background Maxwellian temperature, $m$ the species mass, and $j$ the species identifier. The upper end of the magnetic moment box was set at $9T_{j0}/B_{ref}$, with $B_{ref}$ the reference magnetic field strength (on-axis). Numerical convergence was verified for all grid dimensions, both in the linear and nonlinear simulations. The convergence studies are detailed in the Appendix. Electromagnetic simulations were carried out, but neglecting ${\delta}B_\parallel$, justified by the low $\beta$ of this Ohmic discharge.

\subsection{Synthetic reflectometry diagnostic}
\label{sec:synth}
Details of the synthetic diagnostic are found in Ref.~\cite{hacq16}. To briefly summarize, a 2D full-wave code in the cold plasma limit was employed for propagating the reflectometer wave through the plasma. The experimental $n_e$ and $T_e$ were included as input, together with a $B{\propto}\frac{1}{R}$ magnetic field. In the zone including the resonance, the time-dependent fluctuating density map from the \gene nonlinear simulation was introduced. This corresponds to a toroidal cut of the simulated domain. The full radial extent of the local \gene simulation was included, centred around $\rho_{norm}=0.37$. The probing frequencies in the synthetic diagnostic were set for resonance at that location, for each case run. The reflectometry signal was simulated in a time-dependent fashion by running the full-wave computation at each \gene timepoint. The $E{\times}B$ Doppler shift to the instability phase frequencies were included in the synthetic diagnostic in a post-processing step.

\section{Linear gyrokinetic simulations}
\label{sec:lin}
\begin{figure}[htbp]
	\centering
	\includegraphics[scale=0.75]{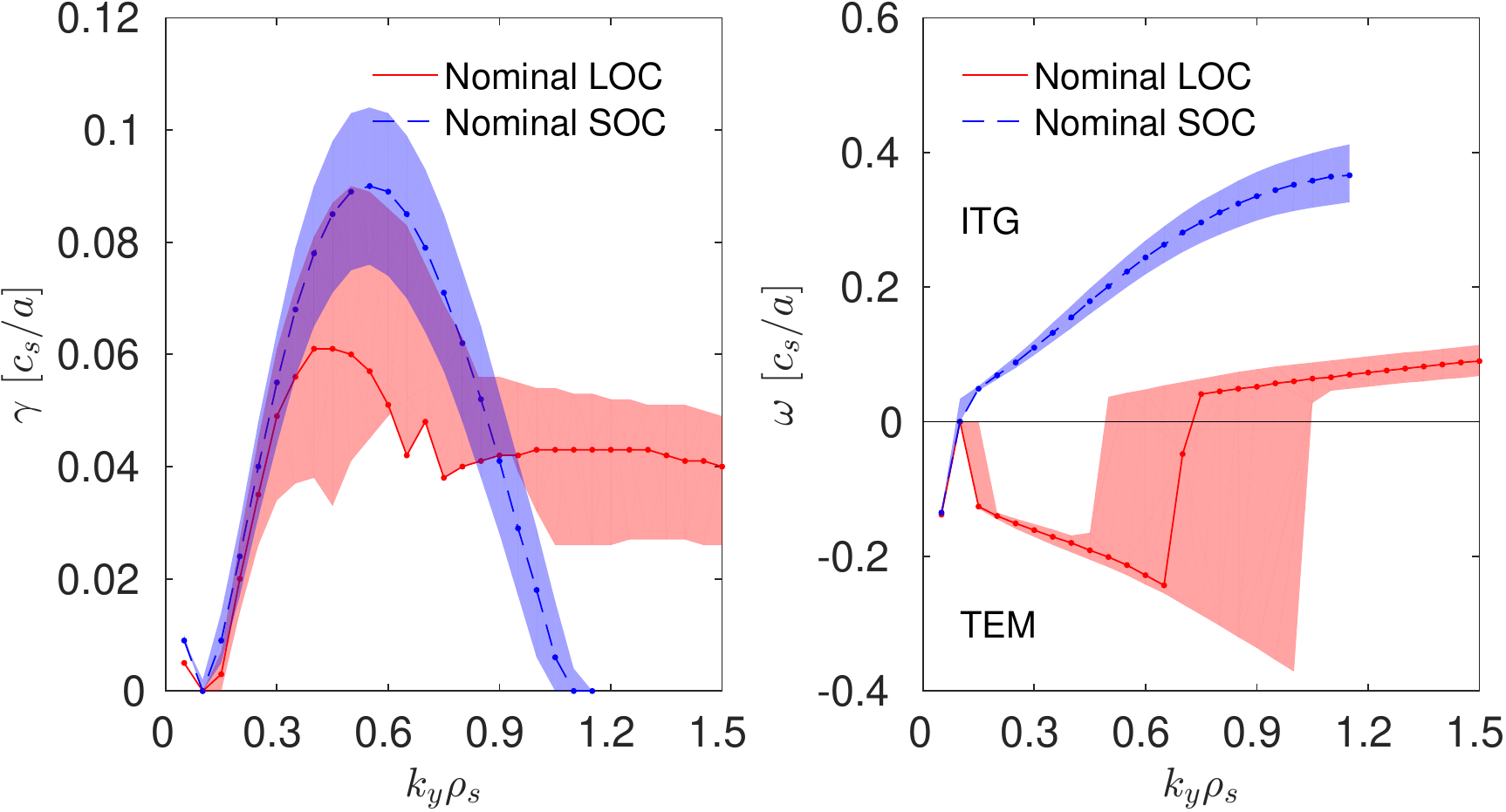}
	\caption{\footnotesize Growth rates (left panel) and frequencies (right panel) of the linear modes in the LOC and SOC phases of Tore-Supra discharge 48102. Positive frequencies correspond to the ion diamagnetic direction. The shaded areas around each nominal curve represent the error propagation of the driving gradient uncertainties}
	\label{fig:figure3}
\end{figure}

The characteristics of the linear modes underlying the turbulent regime in the LOC and SOC phases of the discharge were analysed with 3-species (electrons, deuterium, carbon) linear-\gene simulations. For the nominal parameters, as described in table~\ref{tab:summary}, the dominant linear mode growth rates and frequencies are displayed in figure~\ref{fig:figure3}. The red/blue shaded areas around the LOC/SOC nominal curves correspond to the uncertainties introduced by varying the driving logarithmic gradients $R/L_n$, $R/L_{Te}$ and $R/L_{Ti}$ around their statistical error bars. This variation was carried out by maximizing (and then minimizing) $R/L_n$ and $R/L_{Te}$ together, while simultaneously minimizing (and then maximizing) $R/L_{Ti}$. This procedure leads to the greatest variation in the TEM and ITG instability drives respectively. The logarithmic gradient definitions are: $L_n\equiv-\frac{ns}{\nabla{ns}}$, $L_{Ts}\equiv-\frac{Ts}{\nabla{Ts}}$, where $n$ is density, and $T$ temperature. $s$ is the species identifier, marked as $e$ for electrons, $i$ for main ions (also denoted as $D$ when the identification as deuterium is made explicit), and $C$ for carbon. $R$ is the normalizing major radius.

Since the calculations are with the initial value solver, only the dominant mode for each case is shown. In the SOC phase, the dominant mode is ITG driven, evidenced by the frequency in the ion diamagnetic direction, defined positive in \gene. It remains robustly an ITG mode even when propagating the driving gradient uncertainties. In the LOC phase, the dominant mode is robustly TEM driven in the low wavelength range $0.15<k_y\rho_s<0.45$, which is the range that typically dominates the transport.

The mode with low growth rate at $k_y\rho_s=0.05$ in both the LOC and SOC phases has a microtearing character. However, no tearing parity -- a signature of microtearing modes -- was observed at $k_y\rho_s=0.05$ in the nonlinear simulations. Furthermore, as shown in the Appendix, the nonlinear simulations were converged with respect to increased hyperdiffusion, which decreases this microtearing growth rate. Therefore, it is likely that due to stabilization through nonlinear interactions, this mode has only negligible impact on the saturated nonlinear state. We apply $D_z=10$ in the simulations, where $D_z$ is the \gene parallel hyperdiffusion parameter~\cite{pues10}. No hyperdiffusion in the $x$ and $y$ directions was applied.

\begin{figure}[htbp]
	\centering
	\includegraphics[scale=0.75]{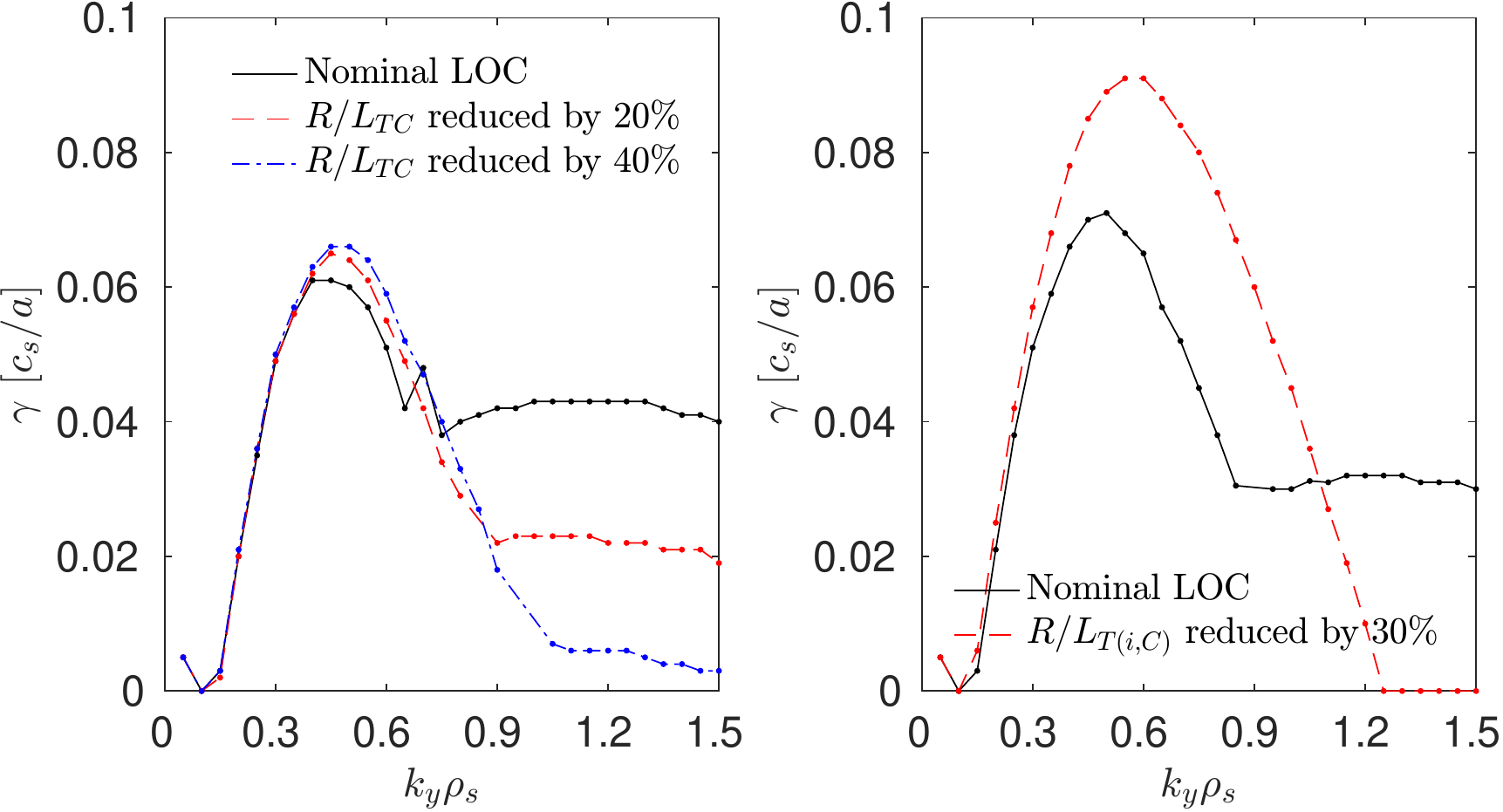}
	\caption{\footnotesize Stabilization of carbon ion temperature gradient modes at $k_y\rho_s>0.7$ in the LOC phase by reduction of carbon ion temperature gradient only (left panel), and by simultaneous reduction of all ion temperature gradients (right panel)}
	\label{fig:figure4}
\end{figure}

The ion mode in the LOC phase at $k_y\rho_s>0.7$, seen in figure~\ref{fig:figure3}, is a carbon driven ITG mode. This is consistent with the significant carbon impurity fraction ($n_C/n_e=0.069$, $n_D/n_e=0.586$) during the LOC, low density phase~\cite{kots95}. The identification of the carbon nature of this mode is clear from a carbon temperature gradient scan shown in figure~\ref{fig:figure4} (left panel), while keeping $R/L_{TD}$ fixed. Reducing $R/L_{TC}$ stabilizes the higher $k_y$ ion mode, while having a negligible effect on the lower $k_y$ TEM. However, due to strong collisional coupling between different ion species, it is inconsistent to vary $R/L_{TC}$ independently from the main ion temperature gradient. Reducing all ion temperature gradients together by 30\% results in a full stabilization of the carbon ITG mode, with a negligible impact on the TEMs in the transport driving range at $k_y\rho_s<0.5$. This is seen in the right panel of figure~\ref{fig:figure4}. The low growth rates and high $k_y$ of the carbon ITG leads to minor impact on the main ion and electron transport levels. Nevertheless, to study a pure TEM regime and reduce uncertainties in interpretation, all nonlinear simulations in the LOC phase reported in section~\ref{sec:nonlin} were carried out with 30\% reduced $R/L_{T(i,C)}$. This was also motivated by the proximity of the nominal LOC case to the TEM-ITG mode transition as seen by the sensitivity study in figure~\ref{fig:figure3}. This proximity may lead to nontrivial nonlinear couplings between TEM and subdominant ITG modes, which we deemed worthwhile to avoid for a pure qualitative comparison between TEM and ITG turbulence regimes.

\begin{figure}[htbp]
	\centering
	\includegraphics[scale=0.75]{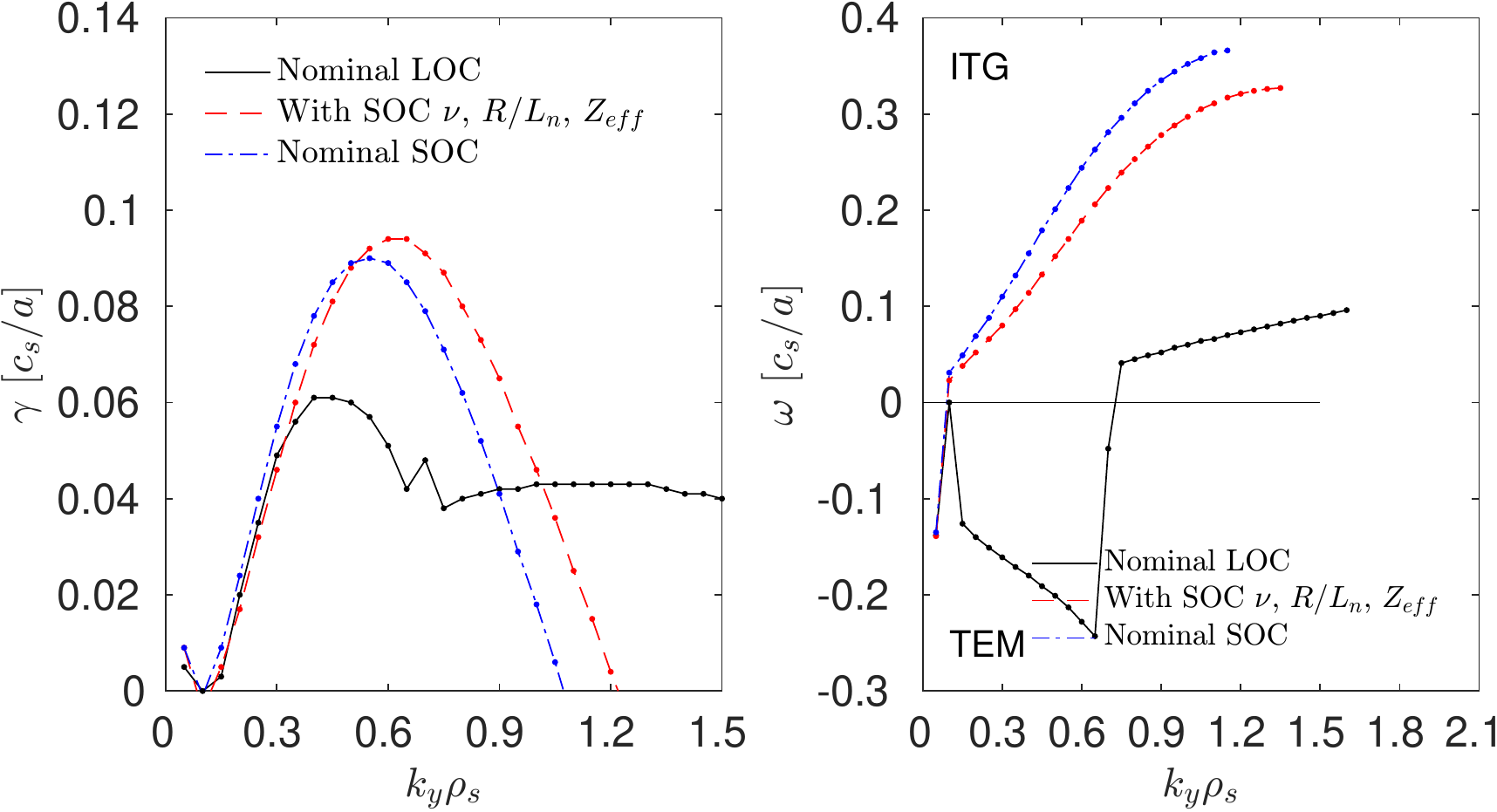}
	\caption{\footnotesize Growth rates (left panel) and frequencies (right panel) for the nominal LOC case (black curve), nominal SOC case (blue curve), and a hybrid case with collisionality, $R/L_n$, and $Z_{eff}$ from the SOC case, and all other parameters from the LOC case}
	\label{fig:figure5}
\end{figure}

The parameters most responsible for the transition from the TEM regime in the LOC phase to the ITG regime in the SOC phase were examined. These were determined to be dilution (represented by $Z_\mathrm{eff}$), collisionality ($\nu$), and the logarithmic density gradient ($R/L_n$). This is displayed in figure~\ref{fig:figure5}. $Z_\mathrm{eff}$, collisionality and $R/L_n$ were replaced in the LOC case to the values from the SOC case. The growth rates then become very close to the nominal SOC case, and the modes revert entirely to ITG. 

\begin{figure}[htbp]
	\centering
	\includegraphics[scale=0.75]{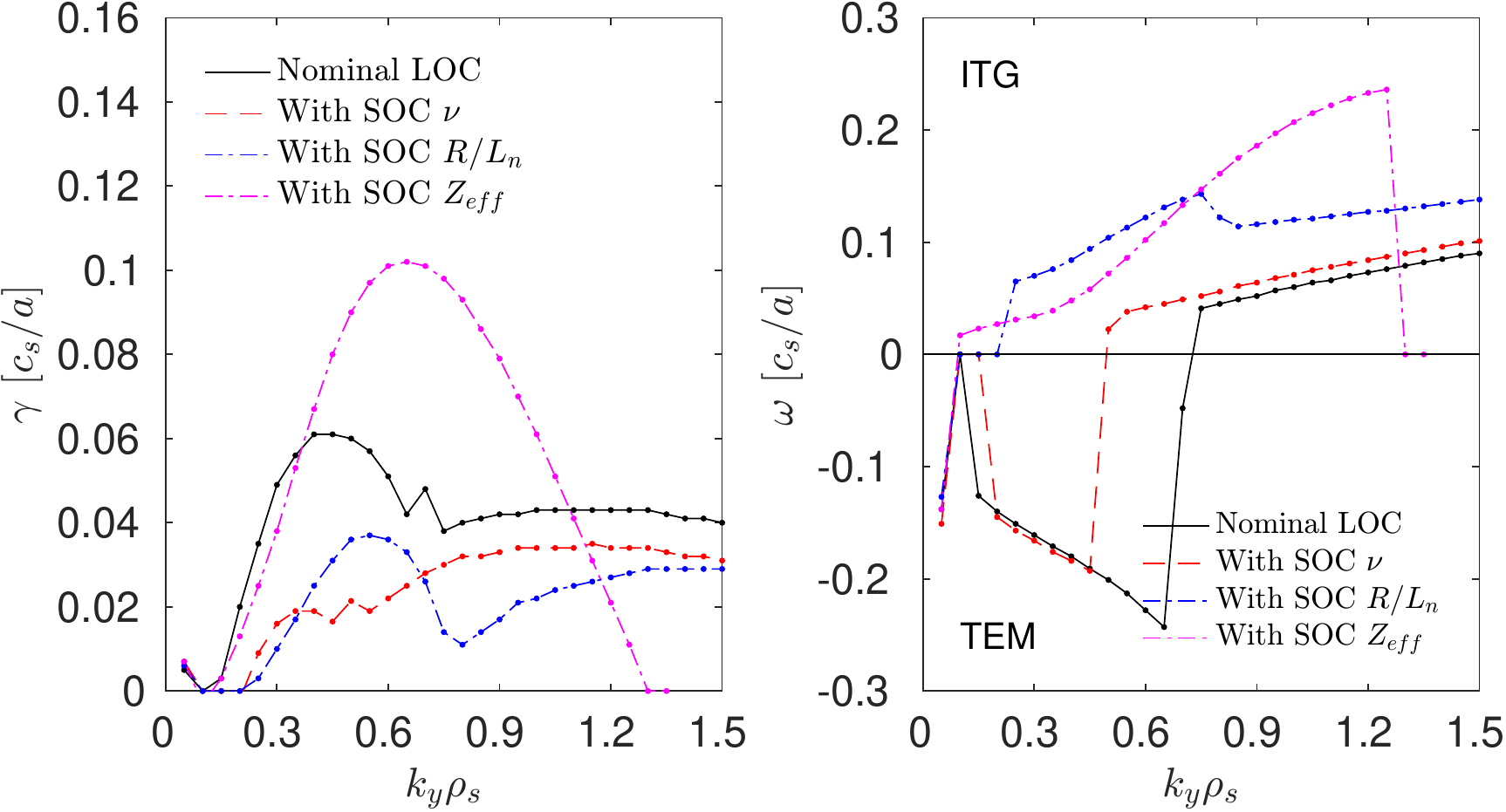}
	\caption{\footnotesize Growth rates (left panel) and frequencies (right panel) for the nominal LOC case (black curve), with LOC parameters apart from SOC collisionality (red curve), with LOC parameters apart from SOC $R/L_n$ (blue curve), and with LOC parameters apart from SOC dilution (magenta curve)}
	\label{fig:figure6}
\end{figure}

Interestingly, out of these parameters, the strongest sensitivity is to dilution. This is seen in figure~\ref{fig:figure6}. While the increase in collisionality significantly weakens the TEM drive, the modes in the transport driving $k_y\rho_s<0.5$ range remain in the electron direction. Decreasing $R/L_n$ does revert all modes to ITG, but with weak growth rates. Decreasing dilution alone leads to the strongest impact with the modes throughout the \kyrhos range reverting to ITG, with similar growth rates to the nominal SOC case. 

\begin{figure}[htbp]
	\centering
	\includegraphics[scale=0.75]{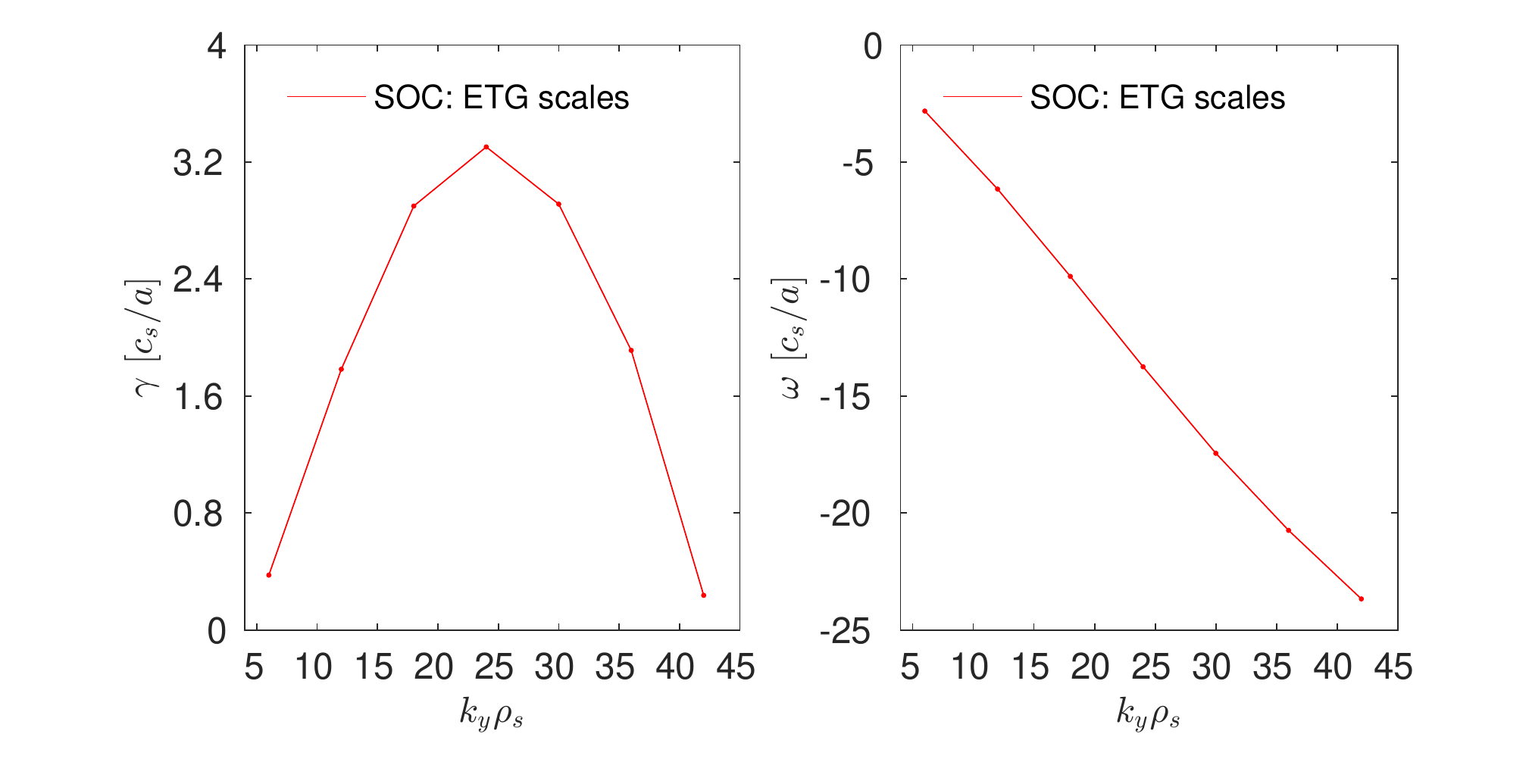}
	\caption{\footnotesize Growth rates (left panel) and frequencies (right panel) for the nominal SOC case on electron scales. The LOC case was stable with respect to ETG modes.}
	\label{fig:figureETG}
\end{figure}

We also carried out a linear analysis of ETG scales. For the LOC case, ETG modes were stable. For the SOC case, ETG modes were weakly unstable. The ETG growth rates and frequencies are shown in figure~\ref{fig:figureETG}. When comparing to the ion-scale modes, we obtain for the ratio of the maximum growth rates in the respective scales $\gamma_{high-k}/\gamma_{low-k}\sim30$. According to the ``rule of thumb'' developed in Ref.~\cite{howa16b}, this ratio likely would lead to negligible ETG fluxes in a multi-scale nonlinear simulation, due to strong interaction with ion-scale eddies. However, we did not have the computational resources to verify this in a dedicated multi-scale simulation for this specific case. 

\begin{figure}[htbp]
	\centering
	\includegraphics[scale=0.75]{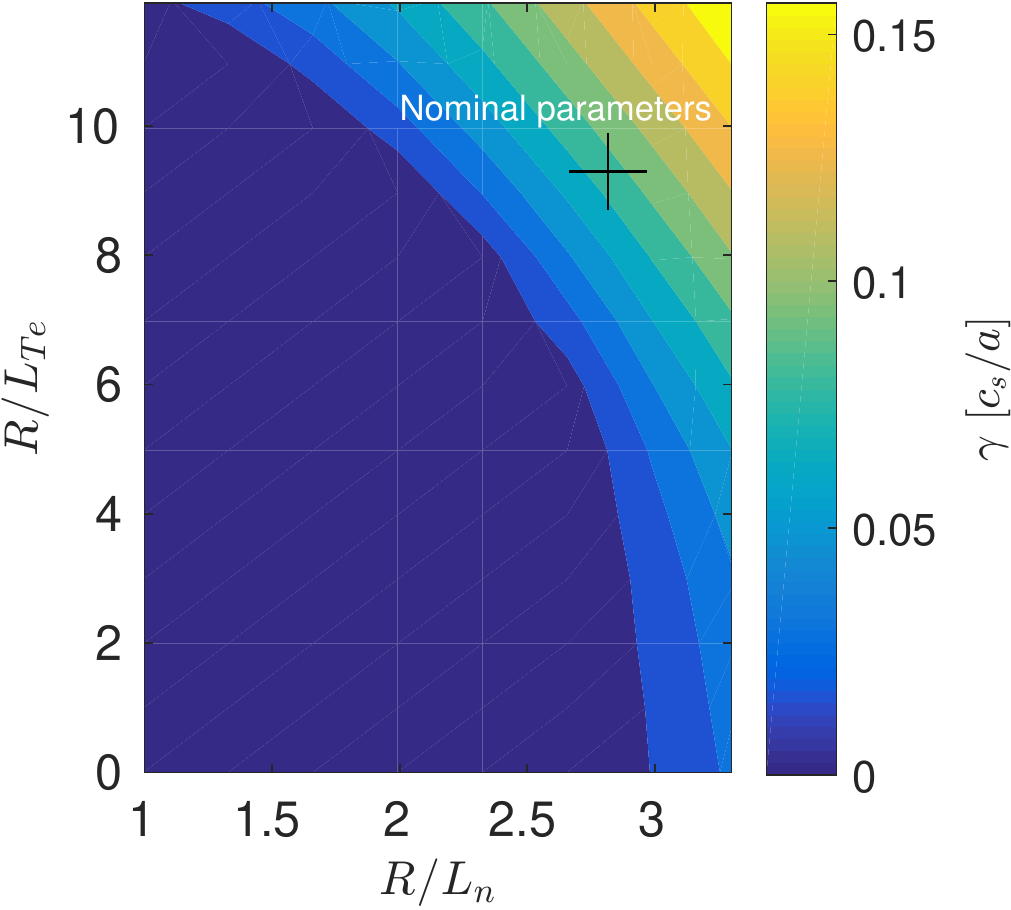}
	\caption{\footnotesize Contour plot of the TEM growth rate as a function of logarithmic density gradient (x-axis), and of logarithmic electron temperature gradient (y-axis). $k_y\rho_s=0.5$ throughout the scan}
	\label{fig:figure7}
\end{figure}

Finally, we investigate the sensitivity of the TEM to the driving $R/L_{Te}$ and $R/L_n$. This is shown in figure~\ref{fig:figure7}. The nominal parameters have $\eta_e\equiv\frac{L_n}{L_T}\approx3.3$. This is associated with the electron temperature gradient driven TEM regime, where zonal flows play a minor role in the nonlinear saturation mechanism~\cite{merz08,erns09}. However, despite having $\eta_e>1$, the mode remains concentrated at low-wavelengths, in contrast to observed transition of TEM to high-wavelength at $\eta_e>1$ for Cyclone-Base-Case parameters with $R/L_{Ti}=0$~\cite{erns09}. It is likely that this observation is regime dependent. Furthermore, in our case both $R/L_{Te}$ and $R/L_n$ have similar weight in driving the instability. This differs from the purely density gradient driven TEMs responsible for QC-TEMs reported in the heated H-mode cases in Ref.~\cite{erns16}, where $R/L_{Te}$ played little role. 

Summarizing, the turbulent regime in the LOC phase is robustly TEM in the transport driving $k_y\rho_s$ range. In the SOC phase, the regime is robustly ITG. The primary parameter variation between the phases that leads to this difference is the dilution of the main ions, which reduces as the density is ramped up from $n_D/n_e\approx0.6$ in the LOC (TEM) phase, to $n_D/n_e=0.9$ in the SOC (ITG) phase. 

\section{Nonlinear gyrokinetic simulations}
\label{sec:nonlin}
Both the LOC (TEM) and the SOC (ITG) cases were studied with 3-species (electrons, deuterium, carbon) nonlinear \gene simulations. The primary goal is to compare the simulated frequency spectra to the measured spectra.
\subsection{Power balance comparison}
\begin{figure}[htbp]
	\centering
	\includegraphics[scale=0.80]{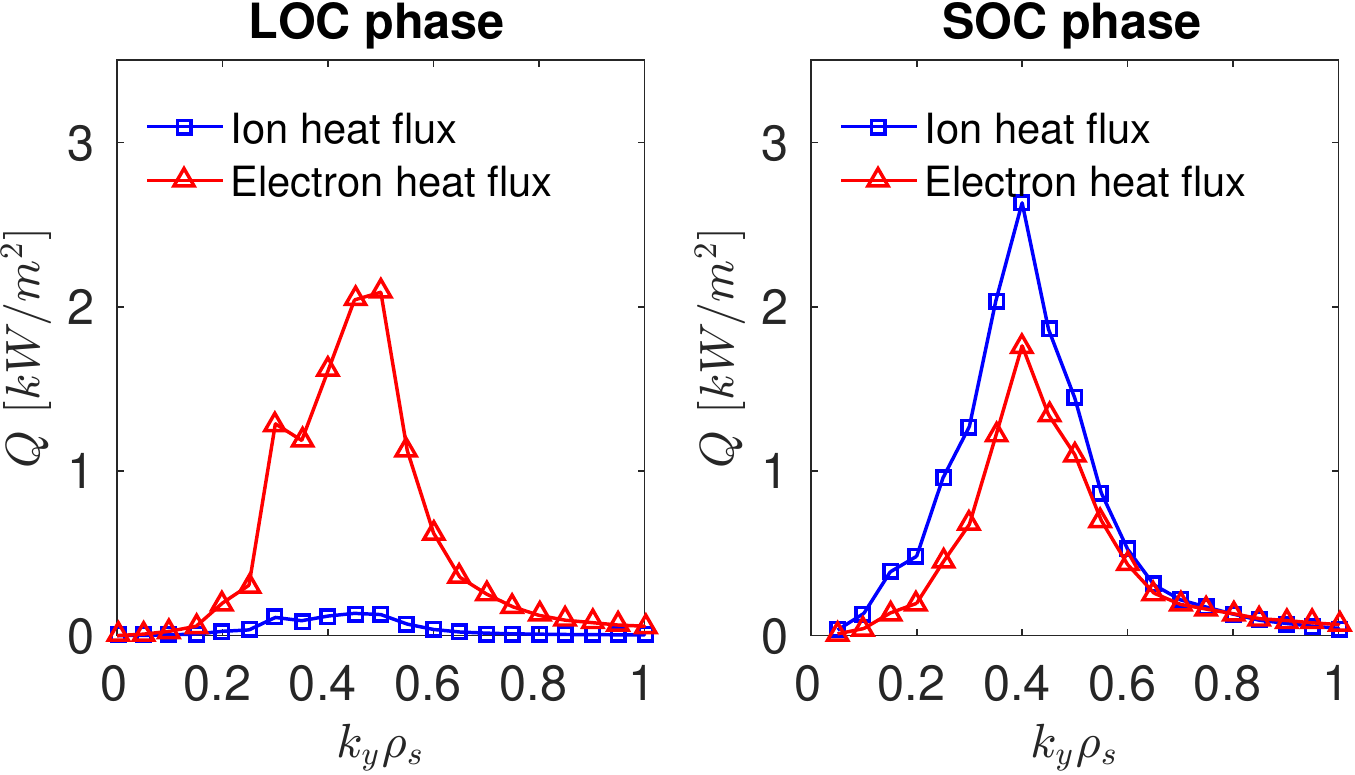}
	\caption{\footnotesize Ion and electron heat flux spectra from nonlinear \gene$ $ simulations for the LOC (TEM) phase (left panel), and for the SOC (ITG) phase (right panel)}
	\label{fig:figure8}
\end{figure}

To examine consistency with the experimental regimes, we first compare the simulated and experimentally estimated ion and electron heat fluxes in both the LOC and SOC phases. The simulated heat flux spectra are displayed in figure~\ref{fig:figure8}. The peak of the flux spectra are downshifted from the peak of the linear growth rate spectra, from $k_y\approx0.6$ down to $k_y\approx0.4$ in both cases. This spectral downshift is a characteristic and ubiquitous phenomenon at both ion and electron scales~\cite{lin05}. The nonlinear LOC simulation is dominated by TEMs, consistent with the linear simulations. This is evident from the propagation of the fluctuations in the electron diamagnetic direction, as seen by the frequency spectra calculated from the nonlinear simulation in the left panel of figure~\ref{fig:figure9}. Furthermore, the simulated heat flux ratio is dominated by electron heat flux, $Q_e/Q_i\approx12$, typical for a TEM dominated regime. The SOC nonlinear simulation is in an ITG regime, consistent with the linear simulations. This is also evidenced by the direction of propagation of the fluctuations, now in the ion diamagnetic direction as seen in the right panel of figure~\ref{fig:figure9}. However, for the SOC nonlinear case heat flux ratio, $Q_e/Q_i\approx0.7$, higher than standard ITG dominated regimes (e.g., see Ref.~\cite{kins06}). This may be indicative that trapped electron effects still have significant influence in setting the fluxes. Table~\ref{tab:heatflux} summarizes the comparison between the simulated heat fluxes, and the experimental power balance as determined by the CRONOS integrated modelling simulation. 

The quoted uncertainties values in Table~\ref{tab:heatflux} include the propagation of $T_i$ and $T_e$ statistical errors into the collisional electron-ion heat transfer for the power balance evaluation, and intermittency in the turbulent heat fluxes in the \gene simulation. The \gene uncertainties are underpredicted, since they do not include the propagation of input parameter statistical and potential systematic errors. Sensitivity studies with respect to a subset of these parameters, not shown here for brevity, show that full power balance agreement in the LOC phase can be easily achieved. However, there is a clear discrepancy between the simulated and experimental power balance in the SOC phase $Q_e$. Since the estimated power balance $Q_e$ value is negative, which is unphysical, this likely indicates that there is an unaccounted for error in the $T_e-T_i$ value, which sets the electron-ion heat exchange. This is particularly relevant for the higher density SOC case, due to the increased collisionality. For both the LOC and SOC phase, a systematic increase of $T_i$, or decrease of $T_e$, would simultaneously decrease $Q_i$ and increase $Q_e$, more so for the SOC phase (higher density) than the LOC phase. This is consistent with the degrees of discrepancy seen in both phases. For the LOC case, the total heat flux $Q_e+Q_i$ agrees between power balance and the gyrokinetic predictions. This is encouraging, since electron-ion heat exchange only redistributes the heat flux between the ion and electron heat transport channels. For the SOC case, there the total simulated heat flux is $\sim50\%$ higher than the power balance total heat flux. This points to an error beyond just the electron-heat heat exchange. However, reducing $T_e-T_i$ closer to 1, would also reduce $T_e/T_i$, stabilizing ITG, reducing the fluxes in the correct direction of the observed discrepancy. While hampered by these uncertainties, the simulated and power balance heat fluxes are consistent with the separation of the phases into TEM electron flux dominated (LOC) and ITG ion heat flux dominated (ITG) regimes. This justifies the comparison of the frequency spectra, and invoking the separate turbulence regimes as a potential source for their observed different characteristics. 

\begin{table*}[tp]
	\small
	\centering
	\caption{\footnotesize Comparison of nonlinear simulated and experimental power balance fluxes. All values in $kW/m^2$. The power balance is calculated from CRONOS integrated modelling simulations, and include propagation of statistical $T_e$ and $T_i$ errors on the ion-electron collisional coupling term. The \gene uncertainties include statistical errors of the quasi-stationary turbulent fluxes, and do not include the propagation of input parameter uncertainties}
	%\vspace{0.15cm}
	\tabcolsep=0.11cm
	\scalebox{0.9}{\begin{tabular}{c|c c|c c}
			\label{tab:heatflux}
			      & \multicolumn{2}{|c|}{Power balance} &  \multicolumn{2}{|c}{\gene} \\
			Phase & $Q_i$ & $Q_e$ & $Q_i$ & $Q_e$  \\
			\hline
			LOC & 4.5$\pm$1.0 & 6.7$\pm$1.0 & 0.73$\pm$0.13 & 10.0$\pm$2.0 \\
			SOC & 14.0$\pm$3.0 & -1.0$\pm$3.0 & 14.0$\pm$2.7 & 9.6$\pm$1.8 \\
		\end{tabular}}
\end{table*}

\subsection{Frequency spectra comparison}

In this section we analyse and compare the simulated frequency spectra from the \gene nonlinear simulations in both the LOC and SOC phases, and compare with a reflectometry synthetic diagnostic. All spectra shown correspond to density fluctuations, which is the measured quantity by the reflectometer. 

\begin{figure}[htbp]
	\centering
	\includegraphics[scale=0.80]{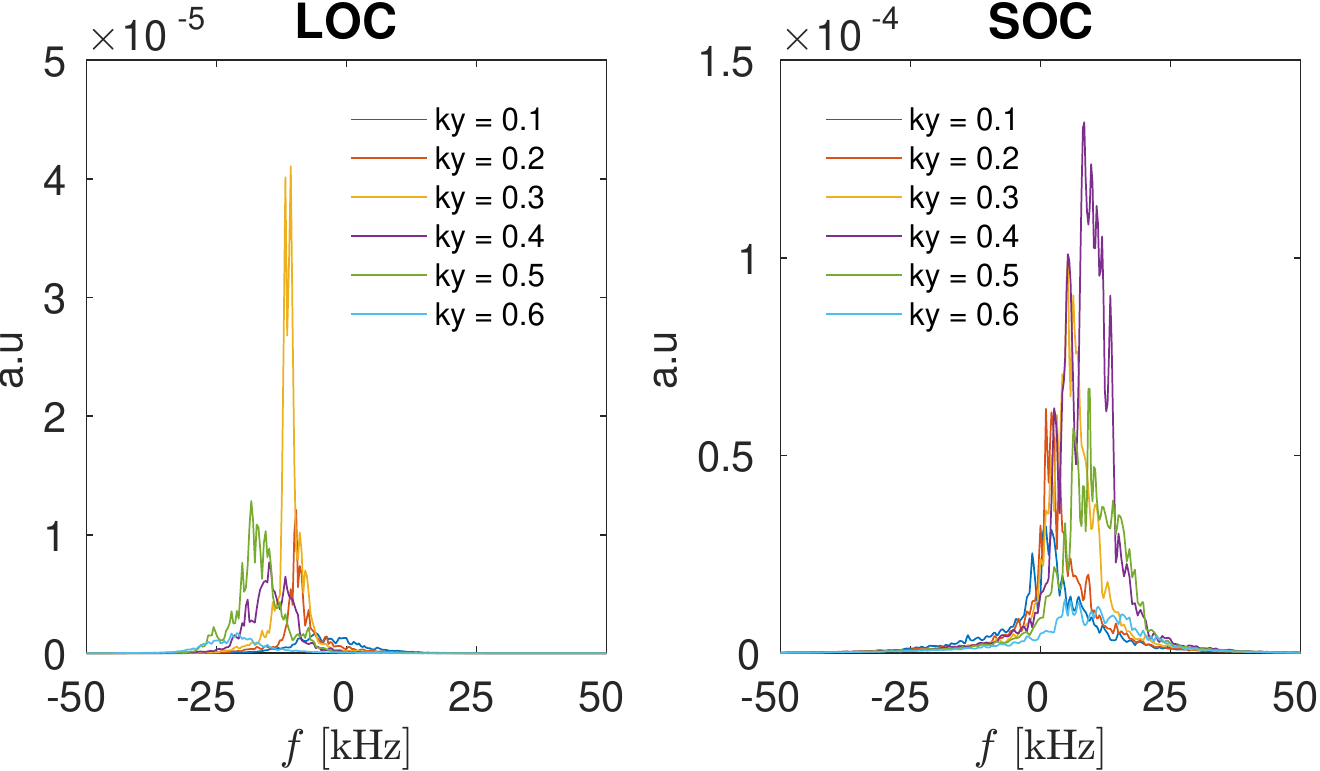}
	\caption{\footnotesize Fourier decomposition (in the binormal coordinate) of the frequency spectra for the LOC phase (left panel) and the SOC phase (right panel) density fluctuations, averaged over $\approx2$~ms during the quasi-stationary saturated state of the nonlinear simulation, corresponding to over $10^2$ turbulence autocorrelation times. $k_y$ is normalised to $1/\rho_s$}
	\label{fig:figure9}
\end{figure}

A $k_y$ decomposition of the density fluctuations is shown in figure~\ref{fig:figure9} for both the LOC and SOC phases. The direct Fast Fourier Transform (FFT) is carried out over multiple (order $10^2$) turbulence autocorrelation times, corresponding to a total of $\sim2$~ms during the quasi-stationary saturated state of each nonlinear simulation. All $k_x$ modes were averaged over for each separate $k_y$. The FFT was carried out at the parallel coordinate corresponding to the outer midplane. For clarity, only a subset of the drift-wave components are plotted in figure~\ref{fig:figure9}. Furthermore, the zonal flow $k_y\rho_s=0$ components are not plotted. In both cases the $k_y=0$ modes are dominated by a $\omega=0$ perturbation, which is not measured by the reflectometer. However, the full spectra, including $k_y=0$, are included as input into the synthetic diagnostic.

\begin{figure}[htbp]
	\centering
	\includegraphics[scale=0.75]{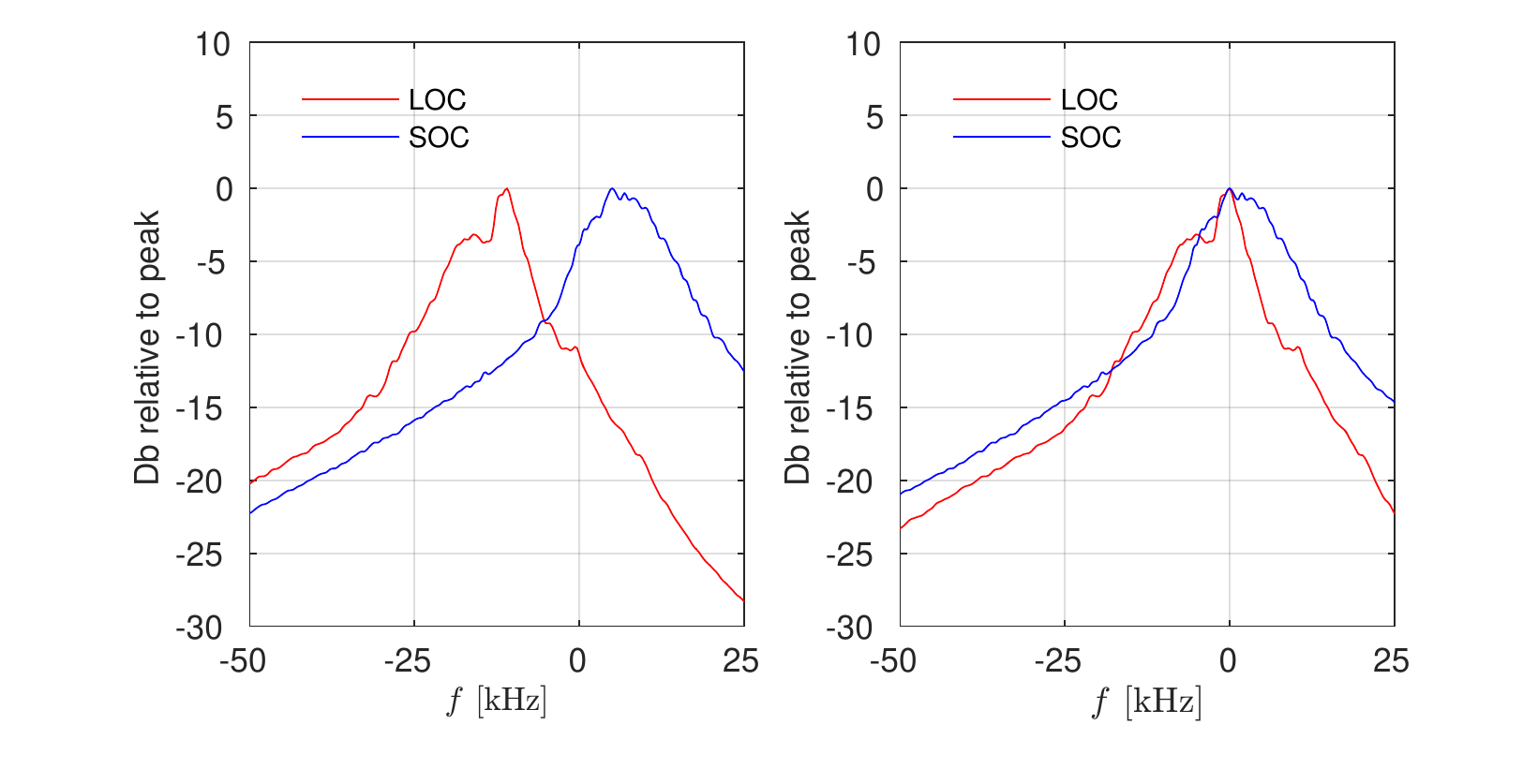}
	\caption{\footnotesize Total integrated frequency spectra for density fluctuations at both the LOC and SOC phases. In the left panel, both cases are plotted with respect to their predicted frequency values. In the right panel, the two spectra are superimposed over each other, centred at their maxima, to illustrate the difference in width. All $k_y>0$ drift-wave components are summed over, and averaged over $\approx1$~ms during the quasi-stationary saturated state of the nonlinear simulation}
	\label{fig:figure10}
\end{figure}

Qualitative differences in the frequency spectra are clearly apparent. In the LOC/SOC phase, the modes propagate in the electron/ion diamagnetic direction, as expected from TEM/ITG turbulence. However, the spectral width of the LOC drift-wave components are narrower than in the SOC phase, as visible in figure~\ref{fig:figure10}. At the $-10~Db$ level, the LOC frequency width is $30\%$ narrower than the SOC frequency width, and a sharp feature at $Db>-5$ is apparent in the LOC case. Furthermore, the TEM fluctuation spectrum has a markedly increased disparity between the peak value ($f\approx-15$~kHz) and $f=0$. This is $\sim10$~Db, compared to the $\sim5$~Db difference between the SOC peak value ($f\approx5$~kHz) and $f=0$. The LOC-case narrow frequency width combined with increased separation between the drift-wave peak and $f=0$ is a key observation in this work, hypothesised to be responsible for TEMs being observable as quasi-coherent-modes, emerging out of the broadband fluctuation spectrum. 

\begin{figure}[htbp]
	\centering
	\includegraphics[scale=0.75]{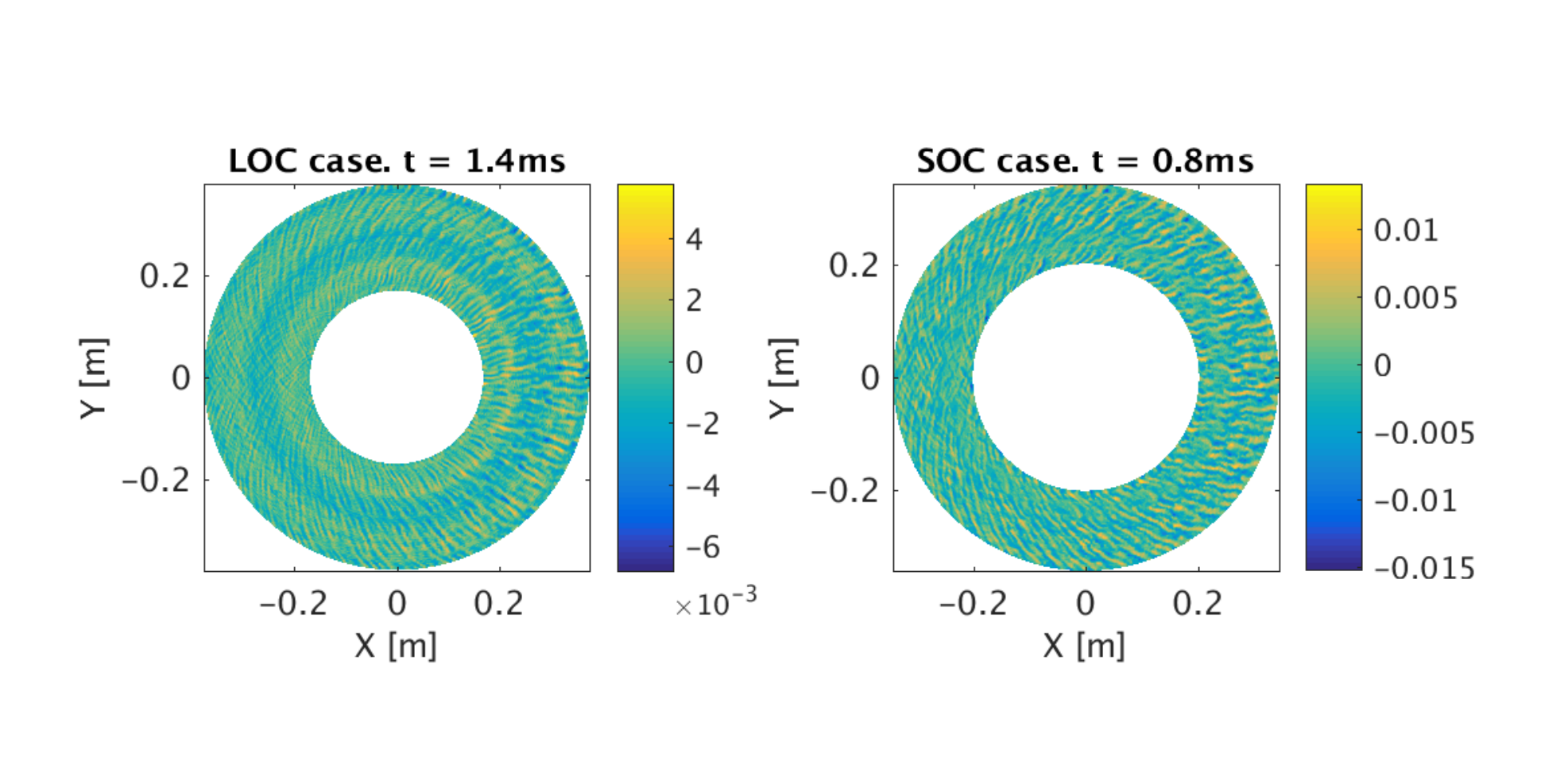}
	\caption{\footnotesize Density fluctuation toroidal cuts from \gene nonlinear simulations of the LOC (left panel) and SOC (right panel) phases. These figures correspond to a single time slice extracted from a series of toroidal cuts corresponding to $\sim2$~ms of simulation time with $\sim1{\mu}s$ resolution}
	\label{fig:figure11}
\end{figure}

The emergence of TEMs from the broadband spectrum is seen in direct comparison with the synthetic reflectometry diagnostic. The reflectometer signal propagates tangentially at constant $\phi$ (toroidal angle). Therefore, the \gene simulation data input for the synthetic diagnostic was constituted from time dependent toroidal cuts of the density fluctuation data, as explained in section~\ref{sec:synth}. The time steps between the toroidal cuts was $2.32{\mu}s$ for the LOC case and $0.74{\mu}s$ for the SOC case. Toroidal cuts corresponding to a single timestep are displayed in figure~\ref{fig:figure11}. 

The measured frequency spectrum of the turbulence is Doppler shifted by the perpendicular $E{\times}B$ velocity of the plasma, i.e. $v_{tot}=v_{phase}+v_{E{\times}B}$. The \gene simulation output contains only the phase velocities. The $E{\times}B$ motion was not self-consistently included in the simulations. For quantitative comparisons with the measurements, the $E{\times}B$ velocity must be taken into account. The Doppler shift was included as a post-processing step of the synthetic diagnostic output. However, this $v_{E{\times}B}=\frac{E_r}{B}$ must be determined. 

We estimate $E_r$ from a neoclassical calculation assuming a ripple dominated regime. The significant magnetic field ripple in Tore Supra justifies this assumption. Furthermore, Ohmic plasmas as in our case have no external torque injection. $E_r$ is set by the ambipolarity constraint for balancing thermal ion losses from ripple wells. This assumption and calculation leads to an estimated $E_r=\frac{T_i}{e}[n_i'/n_i+3.37T_i'/T_i]$, which is validated at Tore Supra by Doppler reflectometry~\cite{trie08}. This radial electric field leads to a velocity in the electron diamagnetic direction. In the synthetic diagnostic output, this Doppler shift of $f_{E{\times}B}=k_y\frac{E_r}{B}/2\pi$ was included in a post-processing step, and included the $1/R$ radial dependence of $B$. Note that no rotation reversal was observed during the LOC-SOC transition in this discharge~\cite{bern15}, as opposed to other cases~\cite{rice11}. This is also consistent with the ripple dominated $E_r$ regime. 

\begin{figure}[htbp]
	\centering
	\includegraphics[scale=0.80]{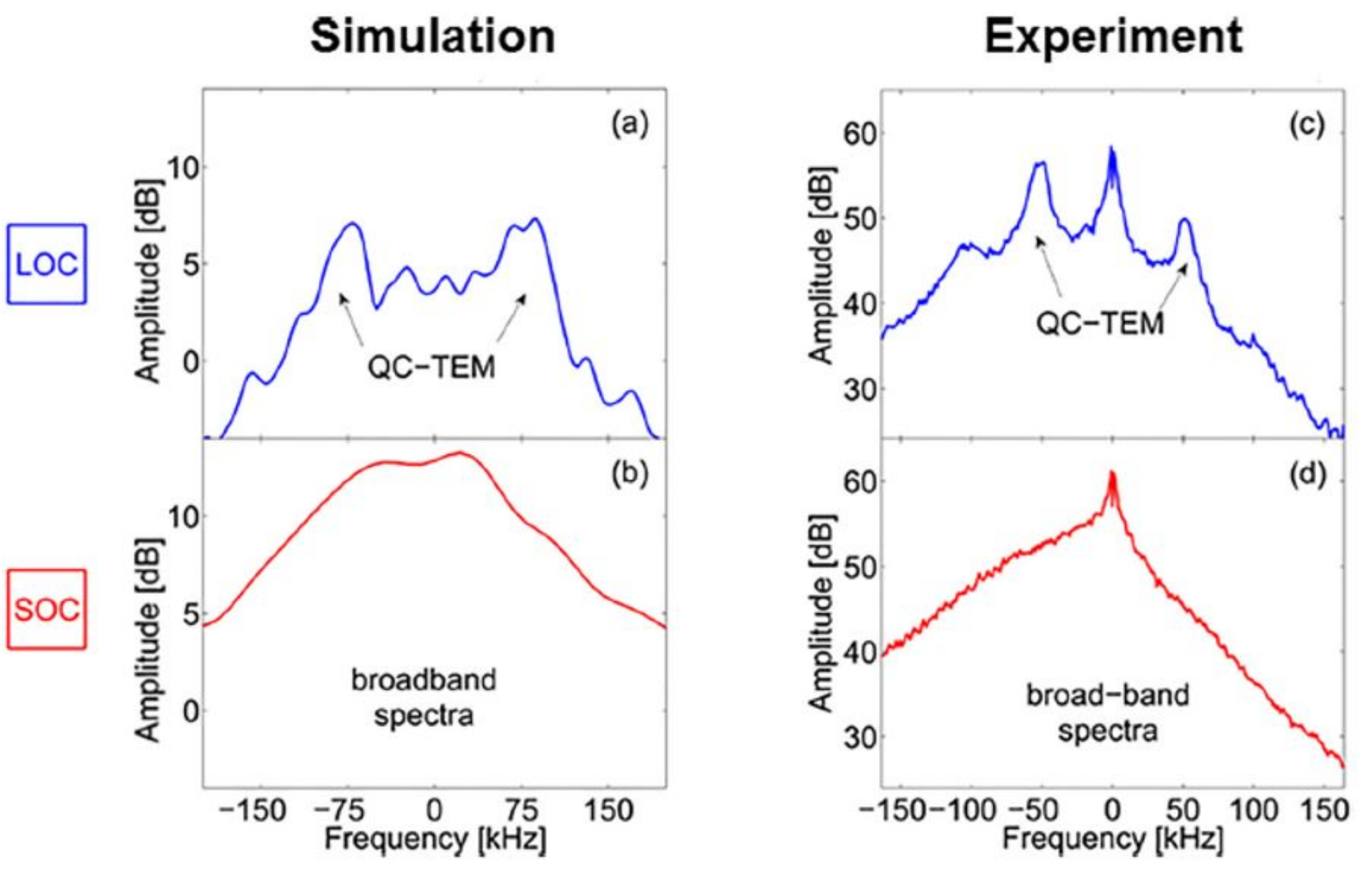}
	\caption{\footnotesize Comparison of the synthetic reflectometry diagnostic for the LOC (a) and SOC (b) phases, to the measured reflectometry signal in the LOC (c) and SOC (d) phases. This figure is reproduced with permission from Ref.~\cite{arni16}. Note that the absolute calibration of the fluctuation amplitude was not carried out for either the measured signal or synthetic diagnostic. However, since the plots display logarithmic units in the y-axis, the difference in scale is apparent as a constant, systematic displacement}
	\label{fig:figure12}
\end{figure}

This shift is of the order $f_{E{\times}B}\approx50$~kHz for $k_y=0.3$ for both SOC and LOC phases (and is of course $k_y$ dependent). This is larger than the drift wave frequencies, which are in the 10-40~kHz range (in absolute units). Since the LOC phase (TEM) fluctuations are already in the electron diamagnetic direction, this Doppler shift affects them more. The shift is an important component in the observability of TEMs as quasi-coherent-modes. The impact of the Doppler shift is seen in figure~\ref{fig:figure12}. The Doppler shift increases the already narrow TEM drift-wave frequency peak up to frequencies further separated from $f=0$, such that the peaks are clearly observable above the broadband. For the SOC case, the Doppler shift is less effective since it has to pass the ITG drift-waves through $f=0$, leading to a reduced separation between the Doppler shifted ITG modes and $f=0$. Furthermore, the ITG drift-wave bulk was already broader to begin with compared to the TEM case. The result is a total broadband ITG frequency spectrum with no significant individual peaks. Note that the central peak of the experimental reflectometry signal is a feature of the carrier wave, and was not included in the synthetic diagnostic. We reiterate that the comparison, while recovering the main characteristics of the observations, should not be considered a one-to-one correspondence, due to the different radial locations of the simulation and experiment, $\rho=0.37$ and $\rho=0.18$ respectively. 

The role of $E{\times}B$ Doppler shifts in leading to observable TEMs has also been highlighted by Ernst \textit{et al}~\cite{erns16}. In the DIII-D H-mode cases discussed therein, the $E{\times}B$ velocity was significantly higher than reported here, due to the beam torque. This led to the reported observation of individual TEMs, separated by the high Doppler shift. In our Ohmic case, due to the more modest Doppler shift, the QC-TEM are observed as a combined bunch. 

\subsection{Differences in frequency broadening}
\begin{figure}[htbp]
	\centering
	\includegraphics[scale=0.80]{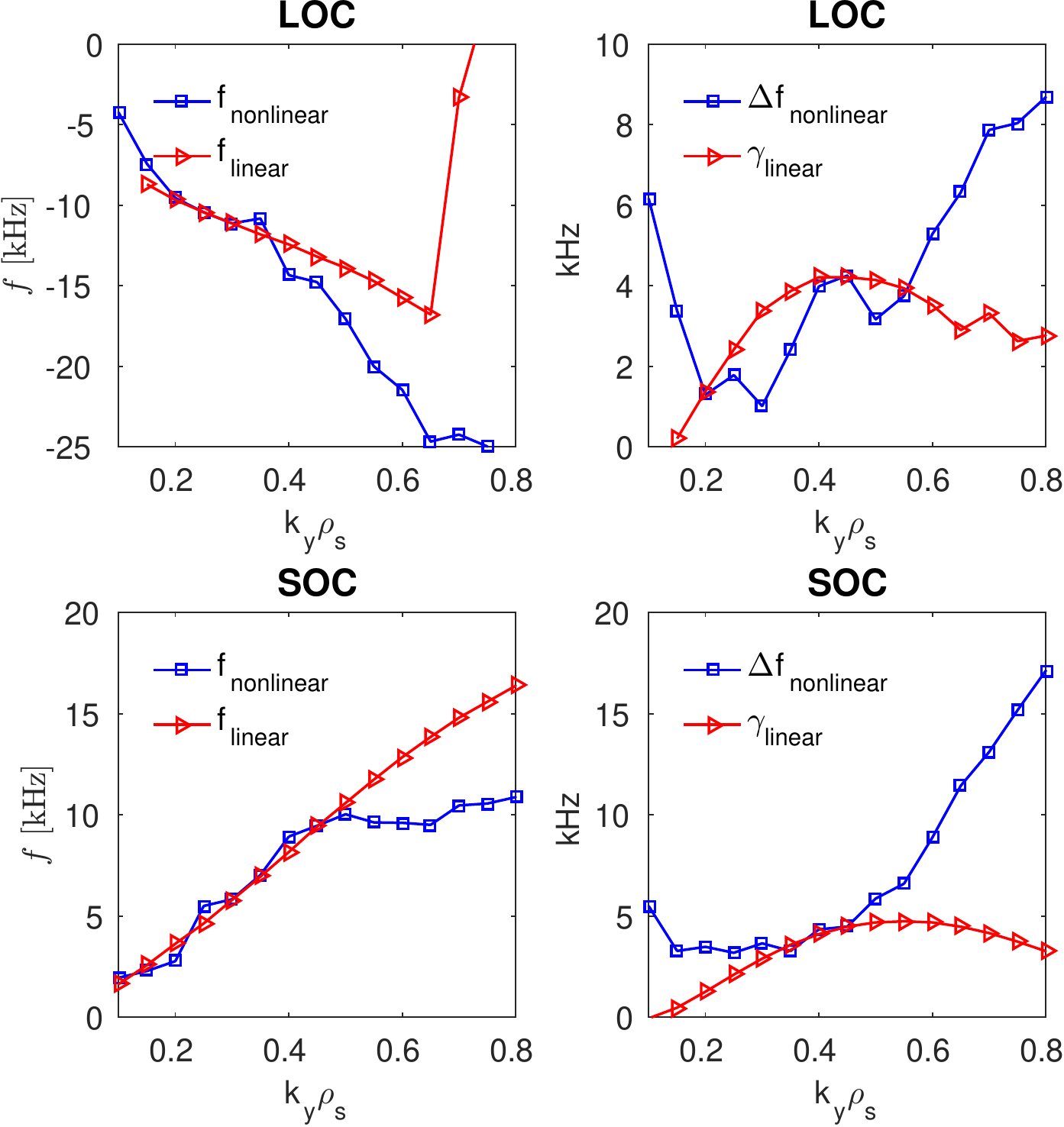}
	\caption{\footnotesize Comparison between linear mode frequency and mean nonlinear frequency peak per $k_y$ for the LOC phase (top left panel) and the SOC phase (bottom left panel). Comparison between linear growth rates and nonlinear frequency broadening for the LOC phase (top right panel) and the SOC phase (bottom right panel). $f_{nonlinear}$ and $\Delta{f_{nonlinear}}$ are the mean and variance of the nonlinear simulation frequency spectrum Gaussian fit per $k_y$. $f_{linear}$ and $\gamma_{linear}$ are the linear frequency and growthrates per $k_y$ for the corresponding linear calculation}
	\label{fig:figure13}
\end{figure}

An interesting open question remains regarding the mechanism behind the observed frequency broadening differences in a temperature gradient driven TEM compared to ITG regime. This difference in quantified in figure~\ref{fig:figure13}, which compares (per $k_y$) the linear and mean nonlinear real frequency, as well as the linear growth rate vs the nonlinear frequency broadening (Gaussian width), for both the LOC and SOC phases. For both phases, the linear and mean nonlinear real frequencies coincide extremely well in the transport dominating $k_y<0.5$ regime. Furthermore, in the peak transport driving region in the SOC case, the frequency broadening coincides with the linear growth rate. This trend is in line with observations in various ITG regimes underlying the formulation of nonlinear saturation rules in quasilinear turbulence models~\cite{bour07,citr12,casa09,casaphd}. However, for the LOC (TEM) phase, the relative frequency broadening is appreciably lower, compared to the SOC (ITG) phase nonlinear frequency broadening. 

Future work should be dedicated to uncovering the source of this fundamental difference. However, we can speculate that this is related to the proposed different nonlinear saturation mechanisms of ITG and TEM turbulence. While ITG turbulence saturates due to coupling with zonal flows, $T_e$ driven TEMs have been observed to saturated via alternative saturation mechanisms~\cite{merz08,erns09}.

To further investigate this, we carried out a $\eta_e\equiv\frac{L_n}{L_T}$ scan in the TEM (LOC) regime. We modified $R/L_{Te}$ and $R/L_n$ from the nominal LOC case, to traverse from $\eta_e=3.3$ to $\eta_e=0.3$ (density driven TEM), while choosing the precise absolute values which maintain a similar level of heat flux to the nominal case. Both linear and nonlinear simulations were carried out. The results are now discussed, comparing the nominal LOC case ($\eta_e=3.3$) with the extreme point of the scan, at $\eta_e=0.3$ (density driven TEM).

\begin{figure}[htbp]
	\centering
	\includegraphics[scale=0.80]{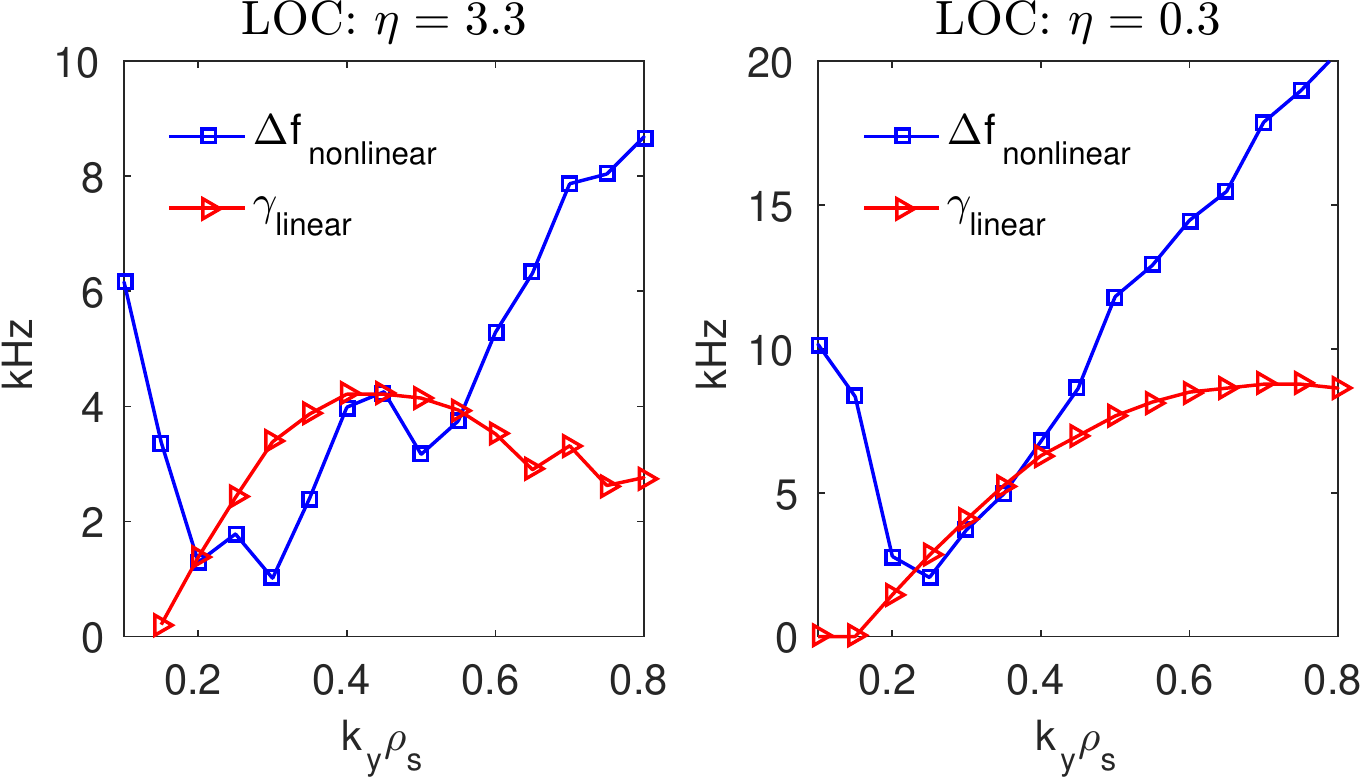}
	\caption{\footnotesize Comparison between linear growth rates and nonlinear frequency broadening for the nominal $\eta_e=3.3$ LOC ($T_e$ driven TEM) phase (left panel), and a density gradient driven $\eta_e=0.3$ TEM case driving the same heat fluxes (right panel)}
	\label{fig:figure14}
\end{figure}

The comparison between the linear growth rates and the frequency broadening in the density driven TEM case ($\eta_e=0.3$) is shown in figure~\ref{fig:figure14}. Compared to the nominal $T_e$ driven case ($\eta_e=3.3$) in the left panel, the density driven TEM case shows significantly increased frequency broadening, and matches the linear growth rates in the transport-driving region, similarly to the ITG regime. Since the frequency broadening comparison in density gradient driven TEMs resembles the ITG case, and since density gradient TEMs are also saturated by zonal flows similarly to ITG~\cite{erns09}, this is suggestive that indeed the difference in frequency broadening may be related to the saturation mechanism. This should motivate further work in this direction. 

\begin{figure}[htbp]
	\centering
	\includegraphics[scale=0.80]{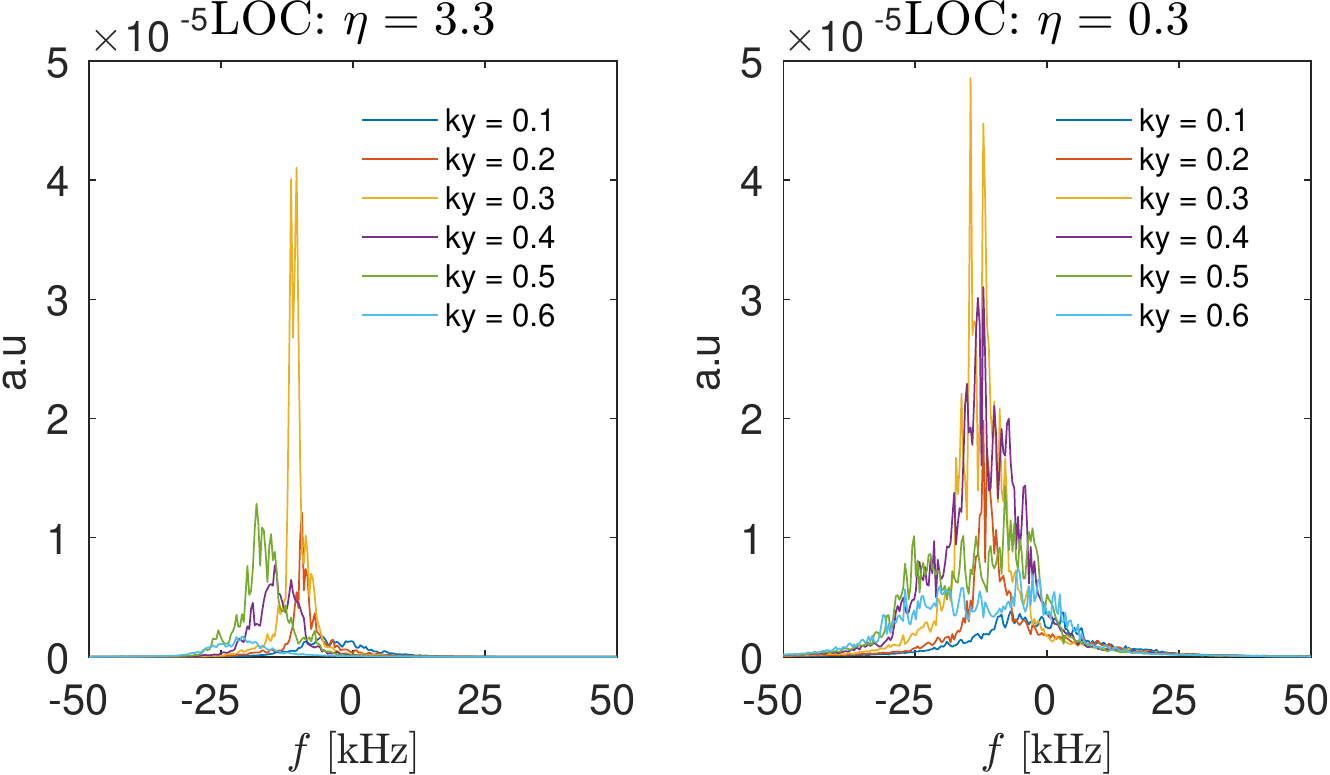}
	\caption{\footnotesize Fourier decomposition (in the binormal coordinate) of the frequency spectra for the nominal $\eta_e=3.3$ LOC ($T_e$ driven TEM) phase (left panel), and the modified density gradient driven $\eta_e=0.3$ TEM case (right panel)}
	\label{fig:figure15}
\end{figure}

\begin{figure}[htbp]
	\centering
	\includegraphics[scale=0.80]{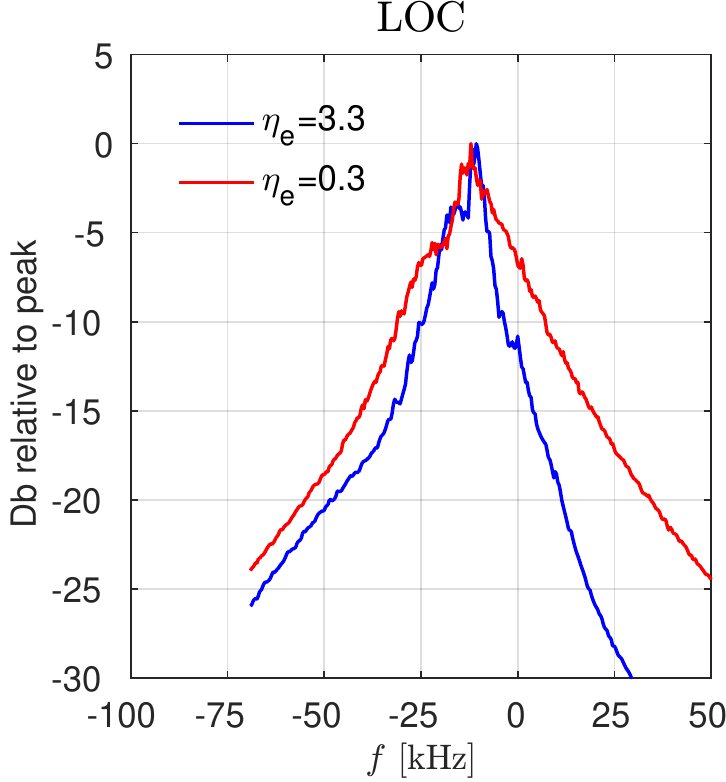}
	\caption{\footnotesize Total frequency spectra for the nominal $\eta_e=3.3$ LOC ($T_e$ driven TEM) phase (blue curve) and a constructed density gradient driven $\eta_e=0.3$ TEM case (right panel)}
	\label{fig:figure16}
\end{figure}

The impact of the increased frequency broadening on the frequency spectrum in the density gradient driven TEM case ($\eta_e=0.3$) is shown in figure~\ref{fig:figure15}. Compared to the nominal $\eta_e=3.3$ case, the $\eta_e=0.3$ density driven case shows more broadband characteristics and less condensation to a small number of modes. The impact on the summed frequency spectrum is shown in figure~\ref{fig:figure16}. The increased broadening of the drift waves in the $\eta_e=0.3$ case leads to a significantly reduced amplitude drop between the drift-wave frequency peak to $f=0$: only 5 Db compared to 10 Db as in the nominal case. This reduced amplitude drop is now similar to the ITG case. This would reduce the observability of the QC-TEM for density gradient driven TEMs.  Interestingly, QC-TEM with density gradient driven TEMs have been observed~\cite{erns16}. However, the $E{\times}B$ shear was significantly higher in those DIII-D H-mode cases compared to the Tore Supra case, further aiding the separation of the modes from $f=0$. It is still unclear whether density gradient driven TEMs are observable as QC-TEMs in a low rotation regime. 

\begin{figure}[htbp]
	\centering
	\includegraphics[scale=0.80]{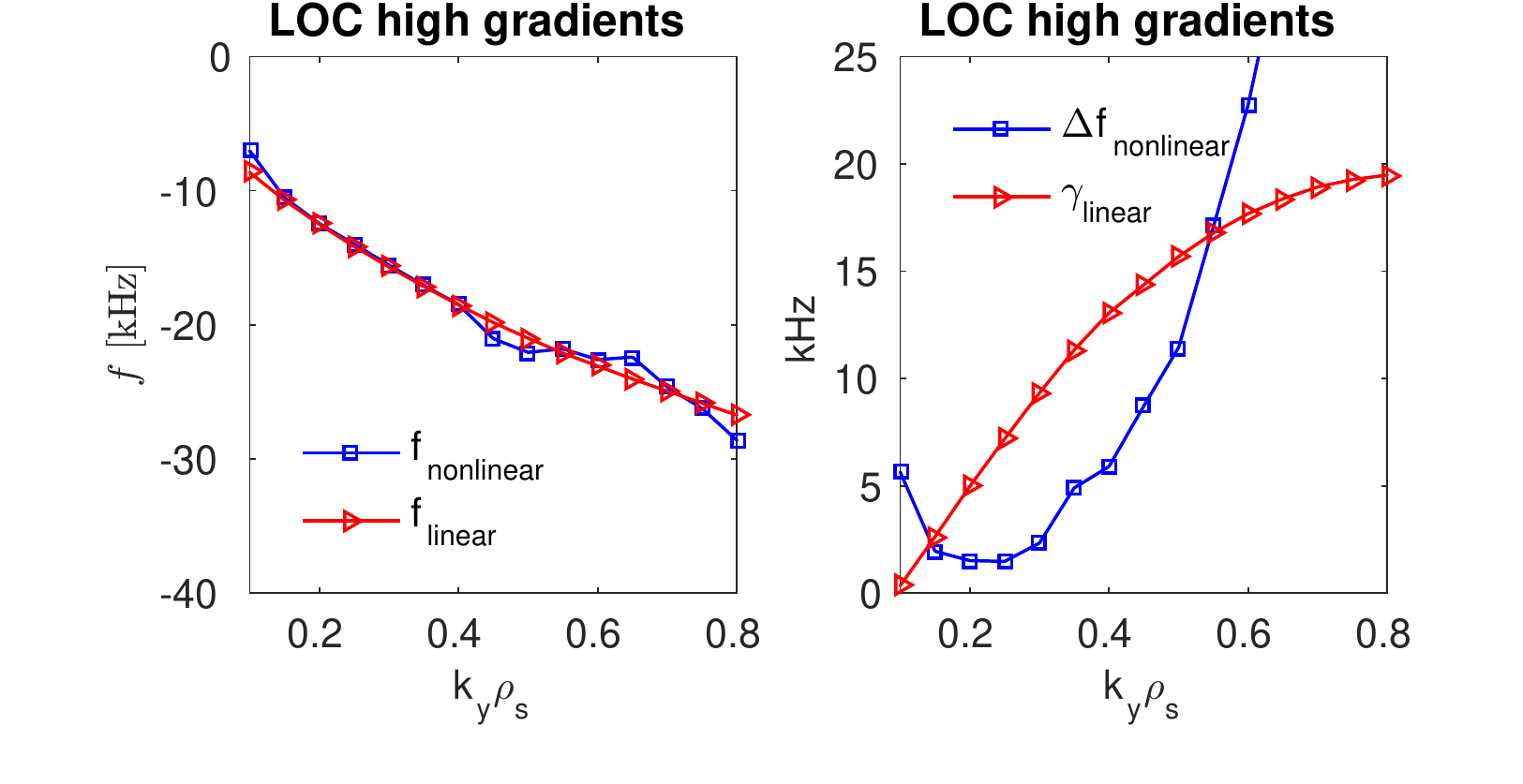}
	\caption{\footnotesize Analysis of nonlinear frequency spectrum compared to the linear spectrum for a high flux LOC-case with $R/L_n$ and $R/L_{Te}$ both increased by a factor 1.4 compared to the nominal case. The left panel shows the comparison between linear mode frequency and mean nonlinear frequency peak per $k_y$. The right panel compares the linear growth rates and nonlinear frequency broadening. $f_{nonlinear}$ and $\Delta{f_{nonlinear}}$ are the mean and variance of the nonlinear simulation frequency spectrum Lorentzian fit per $k_y$. $f_{linear}$ and $\gamma_{linear}$ are the linear frequency and growthrates per $k_y$ for the corresponding linear calculation}
	\label{fig:figureextra2}
\end{figure}

\begin{figure}[htbp]
	\centering
	\includegraphics[scale=0.80]{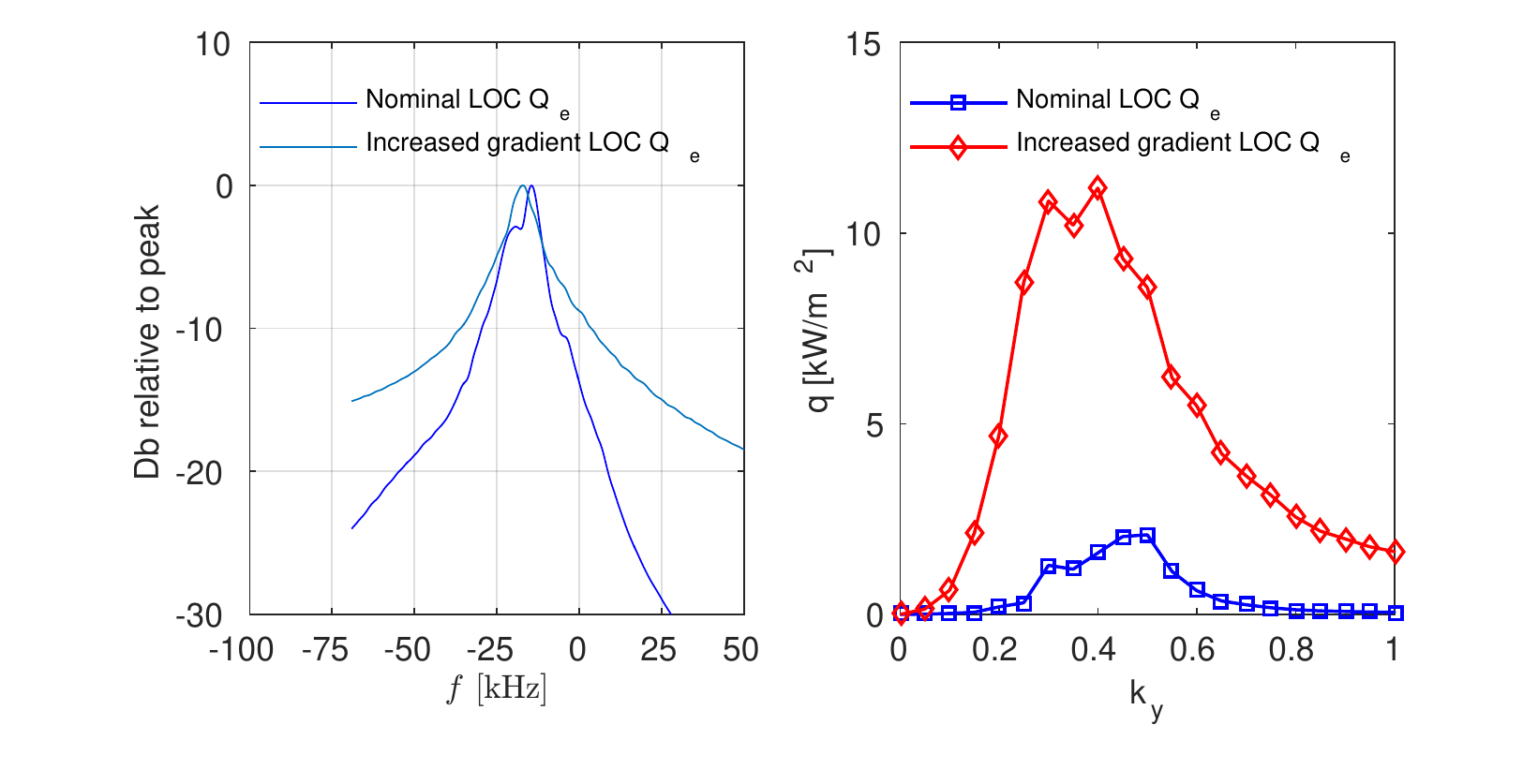}
	\caption{\footnotesize Comparison between the nonlinear frequency and flux spectra between the nominal LOC-case and a high flux LOC-case with $R/L_n$ and $R/L_{Te}$ both increased by a factor 1.4. The integrated frequency spectra are shown in the left panel, where all $k_y>0$ drift-wave components are summed over. The electron heat flux spectra are shown on the right panel. For both analyses, the fluctuations were averaged over $\approx1$~ms during the quasi-stationary saturated state of the nonlinear simulations}
	\label{fig:figureextra}
\end{figure}

Finally, we examine the sensitivity of the nonlinear frequency broadening in the TEM case to the turbulence drive. We carried out a LOC-case nonlinear simulation at $R/L_{Te}$ and $R/L_n$ both increased by a factor 1.4 compared to the nominal LOC-case, (at constant $\eta_e=3.3$). The comparison of the nonlinear and linear frequency spectra of this high gradient case is shown in figure~\ref{fig:figureextra2}. In the left panel, we see a remarkable agreement between the linear mode frequencies and mean-value of the nonlinear frequency spectra. In the right panel, we observe that, as in the nominal (low-gradient) case, the nonlinear frequency widths are significantly more narrow than the linear growth rates in the transport-driving-region of $0.1<k_y\rho_s<0.6$. This is indicative that the narrow frequency broadening is indeed related to $\eta_e>1$ rather than the absolute flux values. The comparison of the integrated frequency spectra of the nominal and high-flux LOC cases are shown in figure~\ref{fig:figureextra}. The right panel compares the electron heat flux spectra, where indeed the increased gradient case predicts increased fluxes by an order of magnitude. Also, due to the increased drive, the linear instabilities arise in a wider spectrum, also evidenced by the wider flux spectrum. In the left panel, the integrated frequency spectra are compared. While the narrow peak in the $Db>-5$ region is similar, the increased gradient case has a significantly wider integrated frequency spectrum at $Db<-5$, likely due to the wider instability range. Without further analysis, we cannot determine if such a feature is observable as QCMs, i.e. if the wider integrated frequency spectrum prohibits the observation, or if conversely other attributes such as increased signal to noise would compensate that. Nevertheless, the salient point is that even at higher fluxes (maintaining the same $\eta_e$ as the nominal case), the nonlinear frequency broadening per $k_y$ remains narrow.

\section{Conclusions}
Quasi-Coherent-Modes appear in measured reflectometry frequency spectra in a Tore-Supra Ohmic plasma LOC phase. These features are qualitatively consistent with the predictions of nonlinear gyrokinetic simulations combined with a synthetic reflectometry diagnostic. The underlying turbulence is identified as $T_e$-gradient TEM driven. Narrow TEM nonlinear frequency broadening, increased separation of the TEM frequency spectrum from zero-frequency ($f=0$), and $E{\times}B$ ripple-well driven plasma velocity leading to Doppler shifts in the electron diamagnetic direction, were all identified as the ingredients allowing this feature to emerge. In the (ITG) SOC phase of the discharge, a broadband measured frequency spectrum was also predicted by the nonlinear simulation and the synthetic diagnostic, consistent with the experimental observations. The robust identification of the frequency spectrum peaks as markers of TEM turbulence may provide valuable realtime information on the turbulence regime of the plasma, important for profile control applications. However, while the onset of QC-TEM in the frequency spectra is linked with unstable TEMs, the inverse does not necessarily hold. A lack of observable QC-TEM does not imply that TEMs are stable. Combined conditions of drift wave frequency spectra and $V_{E{\times}B}$ must be met for this feature to emerge. The main open question is on the mechanism behind the different frequency broadening of the $T_e$-driven TEM and ITG modes. This may be related to the different nonlinear saturation mechanisms, as supported by comparisons with a constructed $n_e$-gradient driven TEM case. Future work focusing on understanding the physics behind the frequency broadening mechanisms is encouraged. This is key to building more robust saturation rules needed for reduced quasilinear turbulence models, and improving their predictive capability.

\section{Acknowledgements}
This work is part of the research programme 'Fellowships for Young Energy Scientists' (YES!) of the Foundation for Fundamental Research on Matter (FOM), which is financially supported by the Netherlands Organisation for Scientific Research (NWO). This work, in partnership with ATEM, is financially supported by the 'Conseil regional Provence-Alpes-C\^{o}te d'Azur'. It has been carried out within the framework of the Erasmus Mundus International Doctoral College in Fusion Science and Engineering (FUSION-DC) and the EUROfusion Consortium. This project has received funding from the European Union's Horizon 2020 research and innovation programme under grant agreement number 633053. The views and opinions expressed herein do not necessarily reflect those of the European Commission. This research used computational resources at the National Research Scientific Computing Center, which is supported by the Office of Science of the U.S. Department of Energy under Contract No. DE-AC02-05CH11231. The authors are grateful to D. R. Mikkelsen for assistance, and to Darin Ernst for fruitful discussions. 

\section*{References}

\bibliographystyle{unsrt}
\bibliography{qcmbib}

\begin{appendix}
	\label{sec:appendix}
\section{Numerical convergence tests}
\subsection{Linear convergence tests}
Numerical convergence with respect to grid resolution, box size, and hyperdiffusion parameter is discussed here, for both the nominal LOC and SOC cases. 

\begin{figure}[htbp]
	\centering
	\includegraphics[scale=0.75]{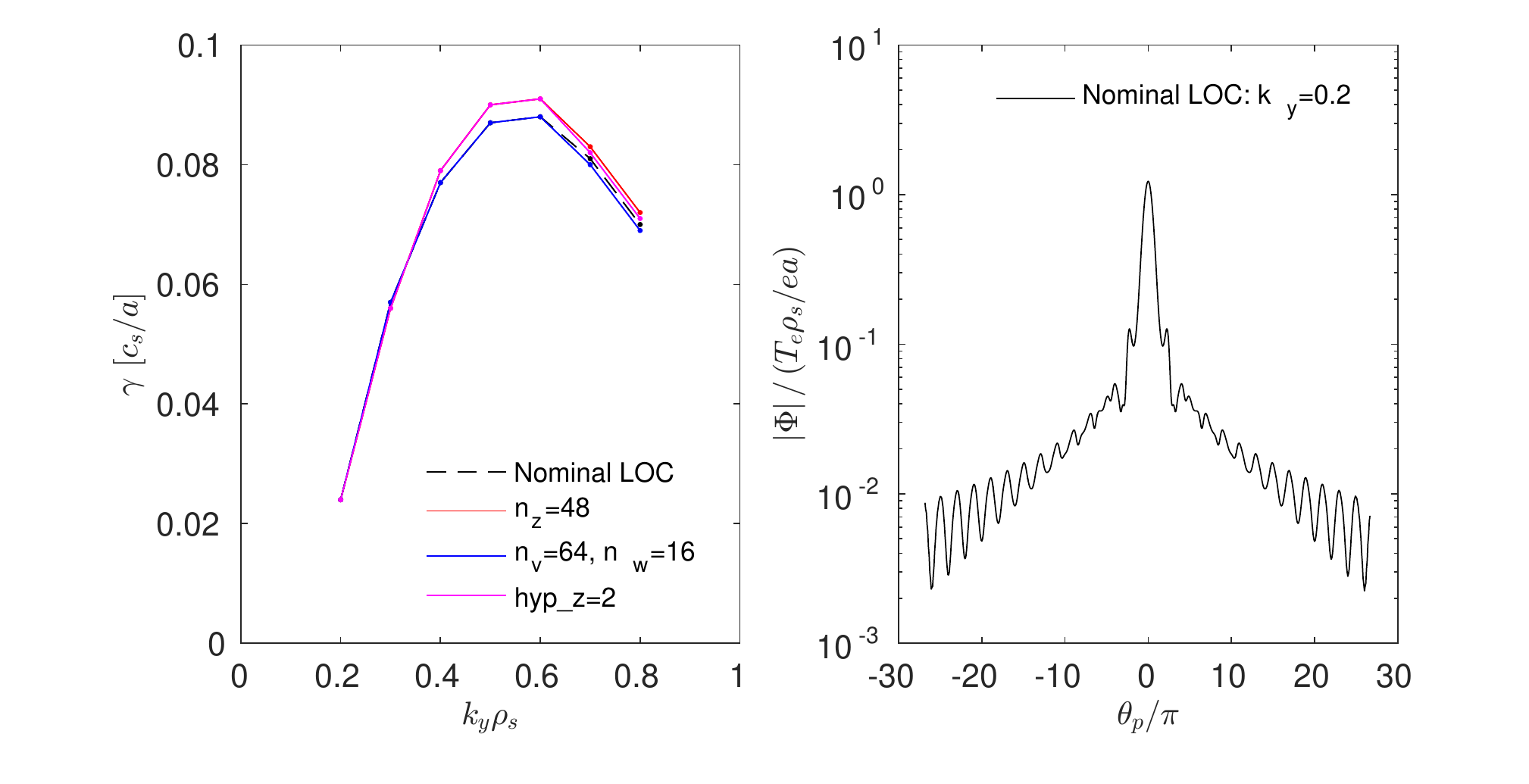}
	\caption{\footnotesize Linear convergence tests for the LOC case. The nominal parameters are $n_z=24$ (parallel direction), $n_v=48$ (parallel velocities), $n_w=12$ (magnetic moments), $D_z=10$ (parallel hyperdiffusion). Growth rate spectra are shown (left panel) comparing the nominal case (black curve) with increased parallel resolution (red curve), velocity-space resolution (blue curve) and reduced parallel hyperdiffusion (magenta curve). Convergence with respect to the number of radial grid points is evident by the strong decay of the eigenfunction amplitude in ballooning space (right panel), shown here for the $k_y\rho_s=0.2$ case}
	\label{fig:figureapp1}
\end{figure}

\begin{figure}[htbp]
	\centering
	\includegraphics[scale=0.75]{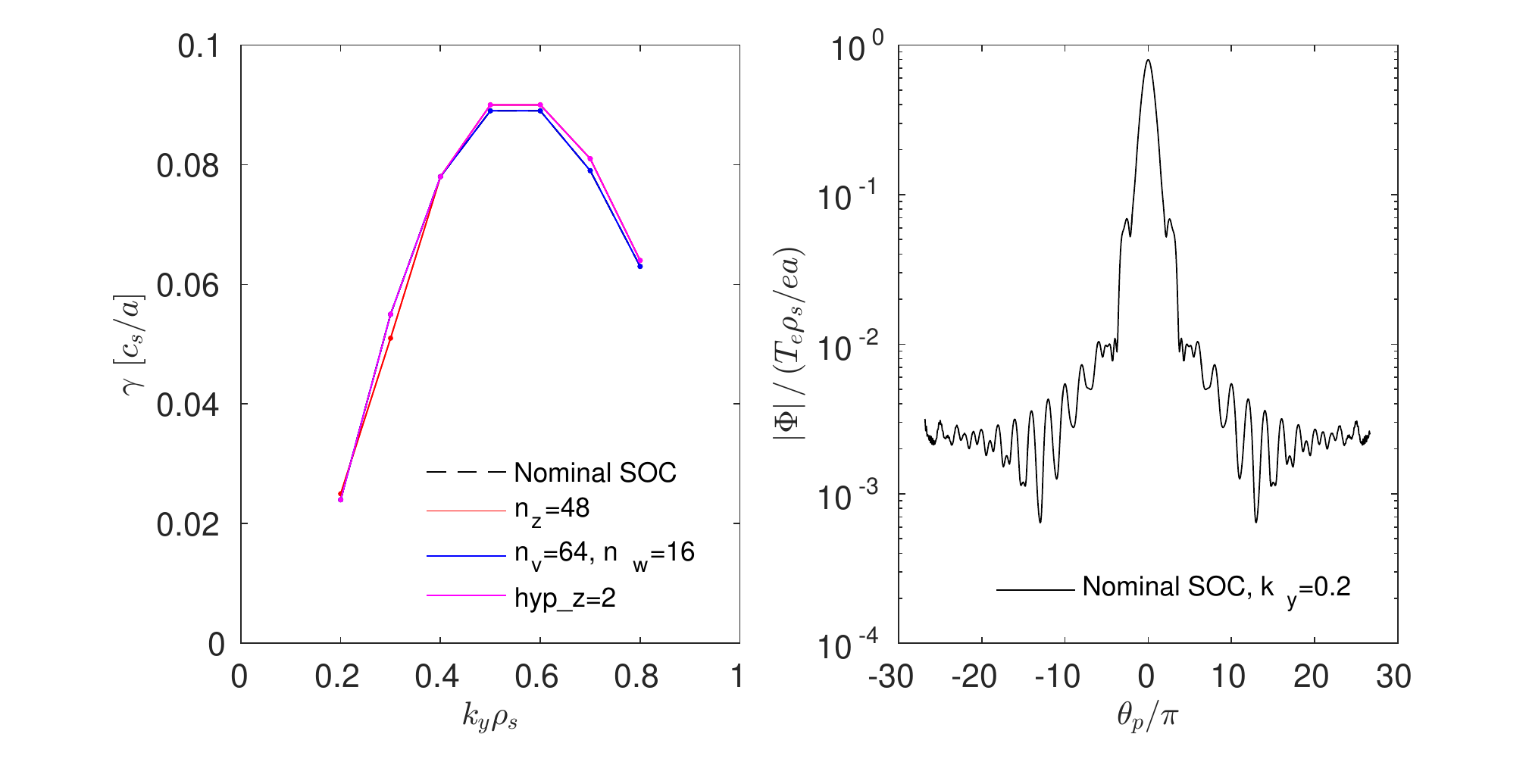}
	\caption{\footnotesize Linear convergence tests for the SOC case. The nominal parameters are $n_z=24$ (parallel direction), $n_v=48$ (parallel velocities), $n_w=12$ (magnetic moments), $D_z=10$ (parallel hyperdiffusion). Growth rate spectra are shown (left panel) comparing the nominal case (black curve) with increased parallel resolution (red curve), velocity-space resolution (blue curve) and reduced parallel hyperdiffusion (magenta curve). Convergence with respect to the number of radial grid points is evident by the strong decay of the eigenfunction amplitude in ballooning space (right panel), shown here for the $k_y\rho_s=0.2$ case}
	\label{fig:figureapp2}
\end{figure}

In figure~\ref{fig:figureapp1} we show the linear convergence for the LOC case, for a spectrum of 8 driftwaves evenly spaced between the range $k_y\rho_s=[0.1~0.8]$. The nominal parameters are 31 radial grid points with $\Delta{k_x}=2\pi\hat{s}k_y$, 24 point discretisation in the parallel direction ($n_z$), 48 points in the parallel velocity direction ($n_v$) and 12 magnetic moments ($n_w$). The nominal parallel hyperdiffusion parameter was $D_z=10$, which was necessary to quench a low-growth-rate numerical instability with tearing parity at $k_y=0.05$. We stress that even without stabilizing this numerical instability, the impact on the nonlinear simulations is negligible. As seen in the left panel of figure~\ref{fig:figureapp1}, insignificant modifications to the growth rates are observed at increased $n_z$, increased velocity space resolution, and decreased parallel hyperdiffusion. Convergence with respect to the radial modes is clearly evident in the right panel of figure~\ref{fig:figureapp1}, by the over two orders of magnitude decay of the mode eigenfunction amplitude at large ballooning angle. The eigenfunctions tend to increase in width in ballooning space for lower $k_y$. Therefore, we show convergence for $k_y\rho_s=0.2$, the lowest unstable $k_y$. Thus, $n_x$ convergence (for the linear runs) is assumed also for higher $k_y$ values. For the SOC case, the nominal parameters were identical to the LOC case, and the linear convergence is shown in the same manner as described above. The convergence plots are shown in figure~\ref{fig:figureapp2}. Linear convergence is thus achieved for both cases.

\subsection{Nonlinear convergence tests}
The detailed breakdown of the dedicated simulations carried out to verify the nonlinear numerical convergence with respect to grid resolution is shown in table~\ref{tab:nlinconv}. For both the LOC and SOC cases, we compare the ion and electron heat fluxes of the nominal case, with the heat fluxes from cases with increased resolution in radial modes $n_x$, toroidal modes $n_y$, grid points in the parallel direction $n_z$, and velocity-space grid points ($n_v$ for parallel velocity and $n_w$ for magnetic moments). In addition the binormal box-size $L_y$ was increased, which sets the minimum $k_y$.

A caveat is that the ``nominal cases'' reported here are not identical to the nominal LOC and SOC cases reported throughout the body of the paper. These cases here, for the convergence checks, include an estimate for the toroidal flow shear. The flow shear slightly reduced the fluxes of the SOC case, and quenched the fluxes of the LOC case such that the LOC ``nominal case'' shown here has 10\% higher $R/L_{Te}$ and $R/L_n$ than the LOC case without flow shear, such that a similar level of heat flux is obtained. In the subsequent analysis, it turned out that the non-physical discontinuous $k_x$ shifts induced by the flow shear model perturbed the FFT diagnostic for the frequency spectra. Thus, in the main cases studied in the paper, no flow shear was included in the \gene simulations. However, we did not repeat the entire convergence analysis without flow shear, both due to limitations in available computing time and the relatively weak impact of the flow shear on the output fluxes and amplitude spectra.

From table~\ref{tab:nlinconv}, it is clear that numerical convergence is obtained for both LOC and SOC nominal cases for increased radial resolution, parallel resolution, and velocity-space resolution.

\begin{figure}[htbp]
	\centering
	\includegraphics[scale=0.65]{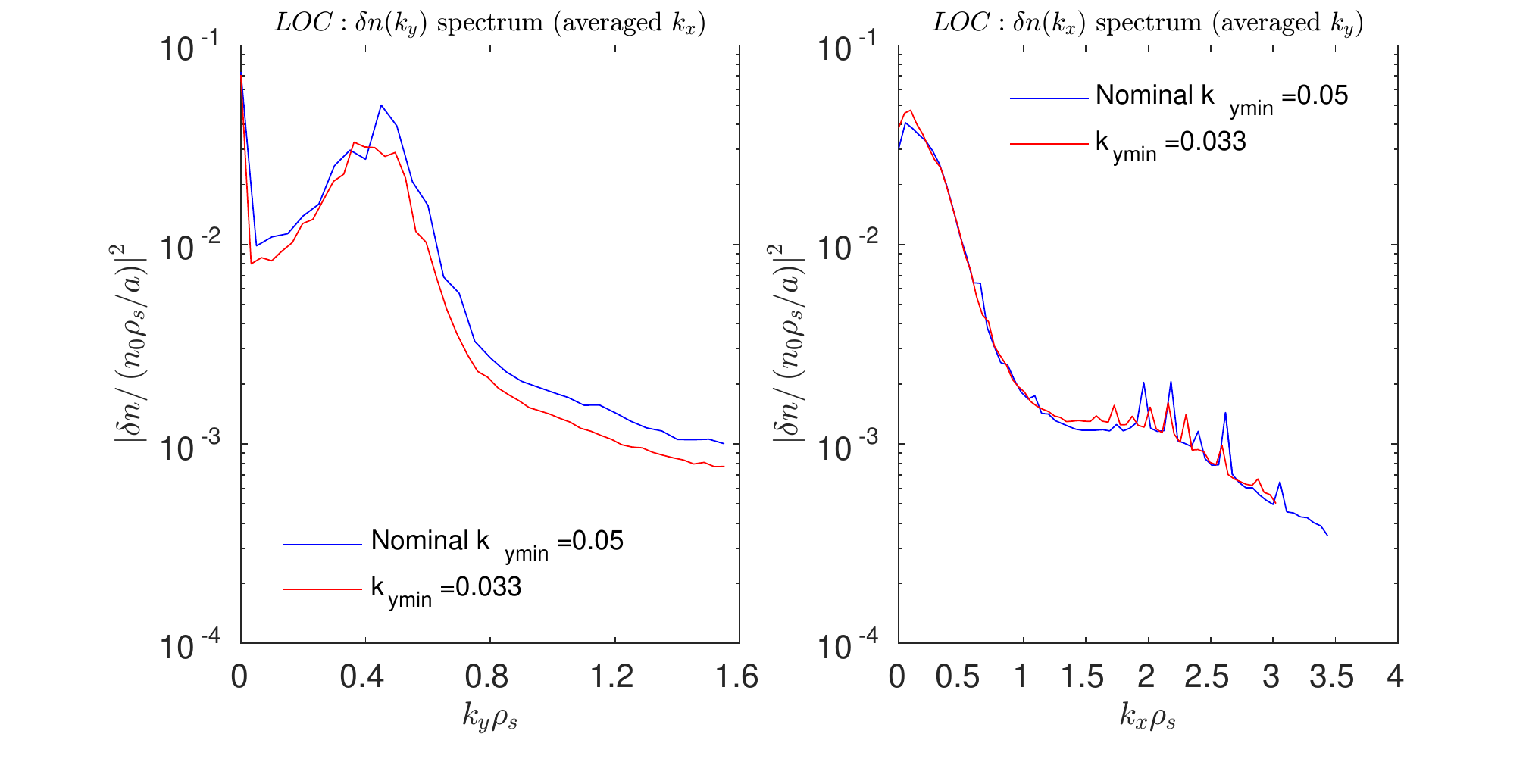}
	\caption{\footnotesize Normalized density fluctuation amplitudes for the nonlinear LOC case, averaged over $\sim1ms$ of the quasi-stationary saturated state. The left panel shows the $k_y$ spectrum, averaged over all $k_x$ modes. The right panel shows the $k_x$ spectrum, averaged over all $k_y$ modes. The spectra are averaged over the parallel direction. The strong decay of the tail at higher wavenumbers is indicative of the resolution convergence in both $x$ and $y$ directions. We compare the nominal case (blue curve) to a case with finer $\Delta{k_y}$ resolution (red curve).}
	\label{fig:figureappspec}
\end{figure}

When increasing the $L_y$ box size (decreased $k_{ymin}$), then the SOC case is clearly converged. However, the LOC case $q_e$ increases by $\sim15\%$ when decreasing $k_{ymin}$ from 0.05 to 0.033. Closer examination of the fluctuation spectra of theses cases, displayed in figure~\ref{fig:figureappspec}, shows that the reason for the higher flux is not a lack of convergence at the box-size $k_y$, but rather due a downshift in the $k_y$ amplitude peak due to the finer $k_y$ resolution. This is consistent with slightly higher fluxes since lower wavenumbers tend to drive more transport. Due to the increased computational expense at $k_{ymin}=0.033$, and the relatively modest difference in flux values and amplitude spectra, it was decided to maintain $k_{ymin}=0.05$ for the main simulations.

When increasing the number of toroidal modes, the eventual flux convergence was at $\sim20\%$ higher values than the nominal case, occurring at $n_y\sim48$ for the SOC case and $n_y\sim64$ for the LOC case. Since the higher $n_y$ simulations show no qualitative difference compared to the nominal case, and the quantitative difference is relatively minor, we deemed $n_y=32$ sufficient for the studies in this paper, to save computing time.

\begin{figure}[htbp]
	\centering
	\includegraphics[scale=0.65]{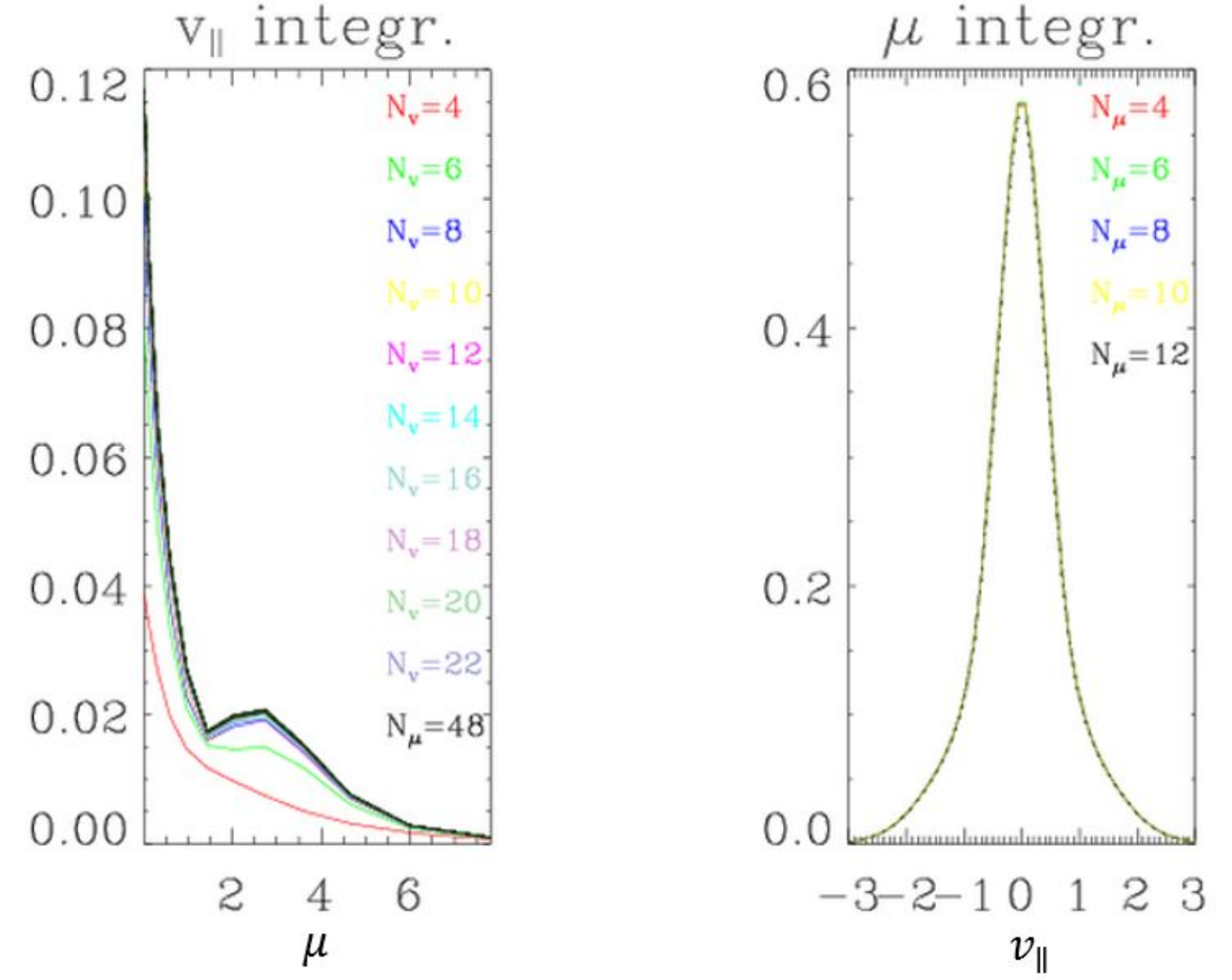}
	\caption{\footnotesize Electron velocity-space convergence check for electrons in the nonlinear LOC case, averaging over $\sim1~ms$ of the quasi-stationary saturated state. The magnetic moment profile $\mu$ is shown in the left panel, integrating over parallel velocity with an increasing number of velocity points until full utilization of the entire (48 point) grid. The parallel velocity profile $v_\parallel$ is shown in the right panel, integrating over magnetic moments with an increasing number of velocity points until full utilization of the entire (12 point) grid. Convergence with respect to the integration is apparent. Convergence with respect to the velocity space box size is also apparent, by the $\sim$2 order of magnitude reduction in amplitude at the upper edge of the boxes.}
	\label{fig:figureapp5}
\end{figure}

Convergence of the velocity-space box sizes is shown in figure~\ref{fig:figureapp5}, for brevity only for the LOC-case electrons (all other cases are similar with respect to the box convergence). The left panel shows the magnetic moment distribution function profile, averaging over all other phase-space coordinates. The parallel velocity integration is carried out also for deliberately reduced resolutions, to test the integration convergence with the nominal $n_v$ value. The right panel is similar, but for the parallel velocity distribution function. It is evident that we achieve convergence with respect to the box-size (as seen by the negligible values at the box edges), and also with respect to the resolution necessary for accurate integration.

To summarize, both linear and nonlinear numerical convergence has been identified for the LOC and SOC cases studied in the paper.

\begin{table*}[tp]
	\small
	\centering
	\caption{\footnotesize Nonlinear convergence study with respect to grid resolution for both the LOC and SOC cases. For all cases, the radial box size $L_x=115$ for LOC, and $L_x=103.7$ for SOC. $D_z=10$ for all cases. Radial box size convergence is evident by the number of eddies captured within the radial extent of the nonlinear simulations shown in Figure~\ref{fig:figure12}.}
	%\vspace{0.15cm}
	\tabcolsep=0.15cm
	
	\scalebox{1.1}{\begin{tabular}{c c | c c c c c c c | c c}
			\label{tab:nlinconv}
			Case & Description & $n_x$ & $n_y$ & $k_{ymin}$ & $n_z$ & $n_v$ & $n_w$ & $D_z$ &  $q_i [W/m^2]$ & $q_e [W/m^2]$ \\
			\hline
			LOC & nominal & 128 & 32 & 0.05 & 24 & 48 & 12 & 10 & 1.77$\pm$0.26 & 16.8$\pm$2.6 \\
			LOC & $n_x$ check & \bf256 & 32 & 0.05 & 24 & 48 & 12 & 10 & 1.65$\pm$0.2 & 17.6$\pm$1.4 \\
			LOC & $n_y$ check & 128 & \bf48 & 0.05 & 24 & 48 & 12 & 10 & 1.95$\pm$0.25 & 19$\pm$2.7 \\
			LOC & $n_y$ check & 128 & \bf64 & 0.05 & 24 & 48 & 12 & 10 & 2.1$\pm$0.3 & 21.0$\pm$3.0 \\
			LOC & $n_y$ check & 128 & \bf96 & 0.05 & 24 & 48 & 12 & 10 & 2.0$\pm$0.26 & 20.1$\pm$2.3 \\
			LOC & $L_y$ check & 128 & \bf48 & \bf0.033 & 24 & 48 & 12 & 10 & 1.95$\pm$0.3 & 19.0$\pm$3.4 \\
			LOC & $n_z$ check & 128 & 32 & 0.05 & \bf48 & 48 & 12 & 10 & 1.6$\pm$0.2 & 17.4$\pm$1.4 \\
			LOC & v-space check & 128 & 32 & 0.05 & 24 & \bf64 & \bf16 & 10 & 1.7$\pm$0.3 & 16.5$\pm$3.0 \\
			LOC & $D_z$ check & 128 & 32 & 0.05 & 24 & 48 & 12 & \bf2 & 1.65$\pm$0.3 & 17.6$\pm$3.2 \\
			\hline
			SOC & nominal & 128 & 32 & 0.05 & 24 & 48 & 12 & 10 & 9.5$\pm$1.4 & 6.4$\pm$1 \\
			SOC & $n_x$ check & \bf256 & 32 & 0.05 & 24 & 48 & 12 & 10 & 9.7$\pm$1.5 & 6.6$\pm$1.0 \\
			SOC & $n_y$ check & 128 & \bf48 & 0.05 & 24 & 48 & 12 & 10 & 11.0$\pm$1.0 & 7.7$\pm$0.7 \\
			SOC & $n_y$ check & 128 & \bf64 & 0.05 & 24 & 48 & 12 & 10 & 11.4$\pm$1.5 & 7.9$\pm$1.1 \\
			SOC & $L_y$ check & 128 & \bf48 & \bf0.033 & 24 & 48 & 12 & 10 & 10.0$\pm$1.1 & 6.7$\pm$0.8 \\
			SOC & $n_z$ check & 128 & 32 & 0.05 & \bf48 & 48 & 12 & 10 & 8.6$\pm$1.2 & 5.7$\pm$0.7 \\
			SOC & v-space check & 128 & 32 & 0.05 & 24 & \bf64 & \bf16 & 10 & 10.4$\pm$1.2 & 7.1$\pm$0.8 \\
			SOC & $D_z$ check & 128 & 32 & 0.05 & 24 & 48 & 12 & \bf2 & 8.6$\pm$1.1 & 5.5$\pm$0.6 \\
		\end{tabular}}
	\end{table*}

\end{appendix}

\end{document}